**Direct growth of mm-size twisted bilayer graphene by plasma-enhanced chemical vapor deposition**

Yen-Chun Chen,[a,b] Wei-Hsiang Lin,[c] Wei-Shiuan Tseng,[a] Chien-Chang Chen,[a] George. R. Rossman,[d] Chii-Dong Chen,[e] Yu-Shu Wu,[b,f] and Nai-Chang Yeh[a,g,*]

[a] Department of Physics, California Institute of Technology (Caltech), Pasadena, CA, 91125, USA

[b] Department of Physics, National Tsing-Hua University, Hsin-Chu 30013, Taiwan, ROC

[c] Department of Applied Physics, California Institute of Technology (Caltech), Pasadena, CA, 91125, USA

[d] Division of Geological and Planetary Science, California Institute of Technology (Caltech), Pasadena, CA, 91125, USA

[e] Institute of Physics, Academia Sinica, Nankang, Taipei 11529, Taiwan, ROC

[f] Department of Electronic Engineer, National Tsing-Hua University, Hsin-Chu 30013, Taiwan, ROC

[g] Kavli Nanoscience Institute, California Institute of Technology (Caltech), Pasadena, CA, 91125, USA

* Corresponding author. E-mail: ncyeh@caltech.edu (Nai-Chang Yeh)

**ABSTRACT:**

Plasma enhanced chemical vapor deposition (PECVD) techniques have been shown to be an efficient method to achieve single-step synthesis of high-quality monolayer graphene (MLG) without the need of active heating. Here we report PECVD-growth of single-crystalline hexagonal bilayer graphene (BLG) flakes and mm-size BLG films with the interlayer twist angle controlled by the growth parameters. The twist angle has been determined by three experimental approaches, including direct measurement of the relative orientation of crystalline edges between two stacked monolayers by scanning electron microscopy, analysis of the twist angle-dependent Raman spectral characteristics, and measurement of the Moiré period with scanning tunneling microscopy. In mm-sized twisted BLG (tBLG) films, the average twist angle can be controlled from 0° to approximately 20°, and the angular spread for a given growth condition can be limited to < 7°. Different work functions between MLG and BLG have been verified by the Kelvin probe force microscopy and ultraviolet photoelectron spectroscopy. Electrical measurements of back-gated field-effect-transistor devices based on small-angle tBLG samples revealed high-quality electric characteristics at 300 K and insulating temperature dependence down to 100 K. This controlled PECVD-growth of tBLG thus provides an efficient approach to investigate the effect of varying Moiré potentials on tBLG.

# 1. Introduction

Graphene, a single layer of carbon atoms arranged in a honeycomb lattice structure with strong in-plane σ ($sp^2$) bonds, has stimulated intense research activities worldwide due to its novel physical properties and great promise for a wide-range of technological applications [1-3]. Graphene samples can be derived by various means, such as micromechanical cleavage of highly ordered pyrolytic graphite (HOPG) [1, 4, 5], chemical exfoliation from bulk graphite [6-8], thermal decomposition or solid-state graphitization of SiC [9-11], chemical reduction of chemically exfoliated graphene oxide [6, 12], and copper-assisted growth by means of thermal chemical vapor deposition (CVD) [13]. The micro-mechanical cleavage method is cumbersome and can only produce small graphene flakes at low yields, and is, therefore, neither efficient for research nor scalable for industrial applications. High-temperature chemical vapor deposition (CVD) system is considered as a preferable way to synthesize uniform and large-size high-quality graphene on Cu at ~ 1000 °C environment and at ~ 600 °C with the aid of plasma [14, 15]. The growth of μm-size to cm-size single crystal hexagonal graphene domains [13, 16] and multi-cm-size graphene films are also achievable by preparing smooth, defect-free substrates to restrict nucleation sites during the high-temperature growth process. However, the multiple steps required in preparing the substrates and the high synthesis temperature in thermal CVD growth of graphene render the process relatively time consuming and expensive, and also incompatible with complementary metal–oxide–semiconductor (CMOS) technology. Recently, we have demonstrated the feasibility of single-step growth of high-mobility monolayer graphene (MLG) by means of plasma-enhanced chemical vapor deposition (PECVD) techniques in a few minutes without the need of active heating [17, 18], which paves ways to a CMOS-compatible approach to graphene synthesis [18].

In addition to MLG, physical properties of few-layer graphene (FLG) systems that are strongly dependent on their stacking arrangements have been actively studied [4, 19-27]. In particular, theoretical predictions [27] and recent experimental verifications [28] of unconventional superconductivity in BLG twisted at a "magic angle" have further stimulated the studies and control of the twist angle in BLG. Early studies of BLG with small twist angles have primarily focused on epitaxial graphene layers grown on SiC [13, 29-31]. However, BLG thus derived generally suffers from a strong substrate interaction such as strong intrinsic n-type doping [13, 29-31]. Alignments between two exfoliated MLG [28], directly exfoliated BLG and folding of MLG [32, 33] are alternative means to achieve different twist angles, although the twist angles thus obtained are generally difficult to control reproducibly.

In this work, we describe our development of an alternative and efficient way of synthesizing BLG with controlled twist angles by PECVD growth on Cu foils in a single step without the need of active heating [17, 18, 34]. Both hexagonal-shaped single crystalline BLG flakes and large uniform BLG films with mm-size are found to be present. These BLG samples are characterized by Raman spectroscopy to investigate the interaction between the electronic and phonon spectra in the BLG system, and to determine the interlayer orientation within the analysis of twist-angle dependent Raman characteristics [35-42]. These studies indicate that most of the angles of twisted BLG (tBLG) tend to arrange in small angle configurations under controlled PECVD growth conditions. Additionally, Kelvin probe force microscopy (KPFM) [43, 44] and ultraviolet photoelectron spectroscopy (UPS) have been taken to investigate the work function of the BLG samples to examine the difference between MLG and BLG due to the shift of Fermi energy with respect to the Dirac point [45]. Scanning tunneling microscopic (STM) images (see Supplementary Information) further reveal periodic Moiré patterns in tBLG systems that are associated with the super-structural length scales when one graphene layer is rotated with respect to the other [27, 46, 47]. Finally, electrical transport properties of the tBLG samples transferred to $SiO_2$/Si substrates have been investigated on more than 20 back-gated two-terminal field-effect transistor (FET) devices made from tBLG samples with small twist angles. We find that all devices reveal the V-shaped conductance (σ) vs. back-gate voltage ($V_g$) dependence and have a mobility value ranging from 4000 to 7000 cm$^2$/V-s at room temperature. Temperature-dependent conductance measurements from 300 K down to 100 K exhibit decreasing conductance with decreasing temperature, suggesting insulating behavior that is consistent with Moiré bands for small-angle tBLGs [27]. Thus, our controlled PECVD-growth of tBLG provides a new and efficient approach to investigate the effect of varying Moiré potentials on tBLG.

This paper is organized as follows. In Section 2, we discuss the synthesis of PECVD-grown single-crystalline BLG flakes and mm-size tBLG films and their characterizations using Raman spectroscopy. We develop a comprehensive database of Raman spectra based on single-crystalline BLG flakes with a range of interlayer twist angles, and then use the database to identify the interlayer twist angle shown in mm-size tBLG films. In Section 3, studies of the work functions of various BLG samples using both KPFM and UPS are detailed. In Section 4, electrical transport measurements on more than 20 back-gated FET devices that are based on BLG samples of small twist angles are described and are shown to exhibit characteristics that are consistent with high-quality material. Finally, we summarize our primary findings and outlook in Section 5.

## 2. PECVD synthesis of graphene and Raman spectroscopic characterizations

*2.1 Experimental setup of the PECVD system and material characterization by Raman spectroscopy*

The experimental setup of PECVD employed to synthesize MLG and BLG samples is consistent with our previous reports [17, 18] and is shown in Fig. S1. The PECVD system is comprised of a microwave plasma source, a growth chamber and a gas delivery system. The plasma source (Opthos Instruments Inc.) includes an Evenson cavity and a power supply (MPG-4) that provides an excitation frequency of 2.45 GHz. The Evenson cavity matches the size of the growth chamber, which primarily consists of a 1/2-inch quartz tube (with the inner and outer diameters being 10.0 mm and 12.5 mm, respectively) and components for vacuum control. The reactant gas delivery system consists of four mass flow controllers (MFCs) for $CH_4$, Ar, $H_2$, and $O_2$. An extra variable leak valve is placed before the $CH_4$-MFC for precise control of the partial pressure of $CH_4$. During the growth process, the pressure of the system is maintained at 25 mTorr. For the PECVD growth substrates, we use 25 μm-thick Cu-foils (Alfa Aesar with purity = 99.9996%) [48, 49]. Prior to the graphene synthesis, the Cu-foils are sonicated in acetone and isopropyl alcohol (IPA) for 5 minutes then dried by nitrogen gas before inserted into the growth chamber. Several pieces of (1.5 × 0.8) $cm^2$ Cu-foils may be first placed on a quartz boat and then introduced into the growth chamber, as shown in the top panel of Fig. S1.

The graphene synthesis on Cu substrates is first preceded by the removal of native copper oxide when the Cu-foils are exposed to $H_2$ plasma at 40 W power for 2 minutes. Planarization of the Cu surface ensues simultaneously within 2 minutes after the oxide removal [50]. The growth chamber is subsequently evacuated until the pressure reaches ~ 25 mTorr, and then a mixture of $CH_4$ and $H_2$ gases with steady partial pressures are introduced, where $H_2$ is used to enhance the dissociation of hydrocarbon [36]. During the initial growth process, graphene nucleates on both the top and bottom sides of the Cu-foils. Continuous exposure of the Cu-foils to plasma at 40 W for 5 ~ 15 minutes in a mixture of $CH_4$ and $H_2$ gases (middle panel of Fig. S1) leads to continuing graphene growth. After the growth, the plasma-heated sample is cooled back to room temperature within at least 30 minutes without breaking the vacuum. The presence of Cu deposits on the quartz tube and holder during PECVD-growth process typically indicates a successful run (see the bottom panel of Fig. S1) because it implies a proper etching rate of Cu surfaces that enables graphene nucleation and growth. Our empirical studies find that 40 W is the optimal plasma power for the growth of both MLG and BLG, and the resulting gas and substrate temperatures under continuous plasma exposure are found to be ~ 425 °C and ~ 400 °C, respectively [17].

Raman spectroscopy with 514 nm laser wavelength and an incident power intensity of 2.5 mW is used to characterize the properties of PECVD-grown graphene, and the typical acquisition time per spectrum is limited to no more than 10 s to prevent overheating the sample by laser. Our choice of Raman spectroscopic characterization is because Raman spectroscopy has been recognized as a practical and non-destructive tool to investigate the interaction between electronic and vibrational spectra exhibited in graphene layered systems. In general, the primary Raman modes associated with graphene characteristics include the doubly degenerate zone center $E_{2g}$-mode associated with the $sp^2$ in-plane phonon vibrations (the G-band at ~ 1584 cm$^{-1}$), the second-order scattering of zone-boundary phonons (the 2D-band at ~ 2700 cm$^{-1}$) [36], and the D (~ 1350 cm$^{-1}$) and D′ (~ 1620 cm$^{-1}$) bands that are characteristics of defect-induced inter- and intra-valley scattering [35, 36, 51-54]. Additionally, the intensity ratio of the 2D- to G-bands, the peak position and the linewidth of the 2D-band are all dependent on whether the sample is MLG, BLG, or tBLG [36]. Specifically, the pronounced π-electron landscapes in bilayer graphene (BLG) system are strongly correlated with their stacking order (*i.e.*, AA- and AB-BLG) and the interlayer twist angles (*i.e.*, tBLG). Therefore, Raman spectroscopic features associated with BLG are generally much richer than those of MLG. For instance, the twist angle (θ) dependent G-band exhibits strong enhancement around the critical angle ($θ_c$ ~ 12°), where the incident laser energy is comparable to the gap between the electron and hole van-Hove singularities (VHS) in the density of states [39], thus resulting in resonant enhancement of the G-band [41, 55], Additionally, interlayer coupling in the BLG gives rise to additional twist-angle dependent Raman modes known as the R- and R′-bands [42,56] as well as a non-dispersive D-band [57].

As we shall show in the following subsections, our twist-angle dependent Raman spectra in tBLG reveal the following general trends: First, for θ < 20°, the G-band intensity ($I_G$) is larger than that of the 2D-band ($I_{2D}$), $I_G > I_{2D}$. Second, the R′-band emerges for 3° < θ < 9°. Third,

Strong G-band enhancement and a tiny R-peak are found at θ ~ θ$_c$ ~ 12°. Fourth, the Raman spectra of tBLG are nearly indistinguishable from that of MLG for θ > 20°, which is attributed to the substantially weakened interlayer coupling in tBLG with θ > 20° [58]. Overall, these Raman spectroscopic characteristics are in good agreement with previous reports [55, 59, 60].

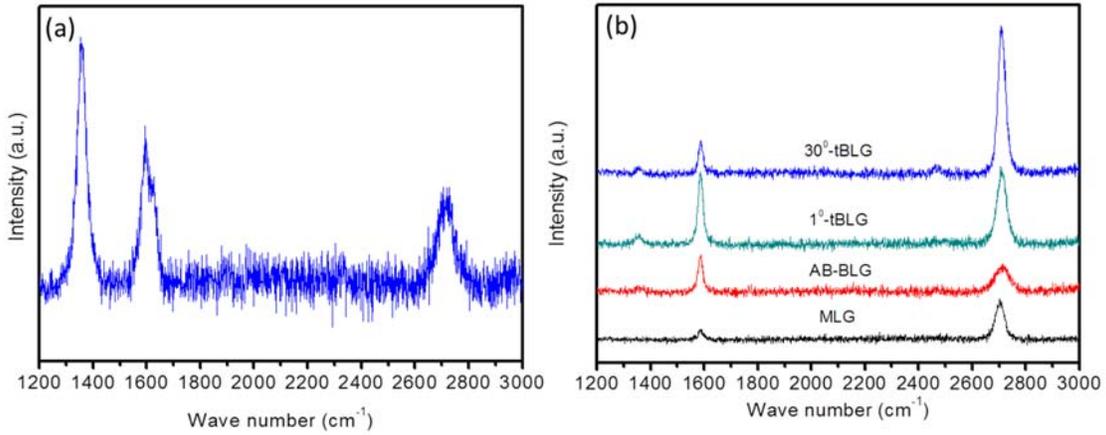

**Fig. 1. (a)** Representative Raman spectrum of graphene grown on the top-side of Cu-foil. The strong intensity of D-band (~ 1350 cm$^{-1}$) and the presence of D′-band (~ 1620 cm$^{-1}$) are indicative disordered graphene, whereas the lower intensity of the 2D-band (~ 2700 cm$^{-1}$) to the G-band (~ 1584 cm$^{-1}$) is consistent of a multilayer graphene sample. **(b)** Raman spectra of MLG, BLG, and tBLG grown on the bottom side of Cu-foil. The Raman spectrum of each curve has been vertically shifted for clarity. The black, red, green and blue curves represent the Raman spectra of MLG, AB-stacking BLG, 1°-tBLG and 30°-tBLG, respectively. The appearance of the D-band at ~ 1350 cm$^{-1}$ shown in the blue curve (30°-tBLG) is attributed to the interlayer interaction.

For graphene grown on the top side of Cu foils, the Raman spectra always reveal D and D′ bands in addition to the G and 2D bands that are associated with pristine graphene [36], as exemplified in Fig. 1(a). The apparent disorder may be attributed to the direct bombardment of energetic particles on graphene grown on the top side of Cu-foils during the PECVD process [17], and the smaller peak intensity of the 2D-band than the G-band suggests the growth of multilayer graphene. In contrast, high-quality MLG and BLG, in both forms of hexagonal single crystals and large films, are found on the bottom side of Cu-foils through proper control of the ratio of CH$_4$ to H$_2$ partial pressures (*i.e.*, P$_{CH4}$/P$_{H2}$), as manifested by the absence of D and D′ bands in the Raman spectra shown in Fig. 1(b).

For the growth of BLG, the twist angles between layers are generally randomly oriented if the PECVD growth parameters are not properly controlled. In the following subsection, we describe our approach to achieve control of the twist angles in the PECVD-grown tBLG.

*2.2 Growth of Bilayer Graphene*

The PECVD-growth processes of BLG are similar to the single-step PECVD-growth of MLG that we developed previously [17] except that the twist angle between the layers may be controlled by the ratio of partial pressures of CH$_4$ and H$_2$, (P$_{CH4}$/P$_{H2}$). Here we divide the growth processes into those for synthesizing single crystalline BLG flakes in Subsection 2.2.1 and those for synthesizing mm-sized uniform BLG films in Subsection 2.2.2. By optimizing the control of P$_{CH4}$/P$_{H2}$, we show that both types of BLG samples display an overall small spread of angular distributions (from AB-stacking to ~ 7.5°), where the interlayer orientation is identified by combined measurements of the hexagonal single crystalline edges between the two layers and Raman spectroscopy.

*2.2.1 Growth of single crystalline bilayer graphene*

Comparing the conditions for growing MLG on Cu, a critical parameter to achieve the growth of BLG on Cu is to adjust P$_{CH4}$/P$_{H2}$ [34]. The rationale for this approach may be understood as follows: the second layer of graphene is known to grow underneath the first layer if excess carbon atoms migrate between the Cu surface and the first layer of graphene, which suggests that a pathway to achieve large-area BLG is to balance the growth rate between the first and second layers of graphene. That is, by slowing the growth rate of the first layer sufficiently to allow excess carbon atoms sufficient time to migrate beneath the first layer to form the second layer [48, 61, 62]. Moreover, during the PECVD process, the growth rate of graphene depends on the rate of etching Cu-foils [50] so that the presence of $H_2$ gas, as well as the resulting radicals, are not only the catalysts necessary to decompose $CH_4$ into C- and H-atoms for ensuing graphene growth but also the agent that controls the etching rate of both graphene [51] and the Cu-foil [17]. Therefore, by adjusting the partial pressure of $CH_4$, $P_{CH4}$, and by keeping $H_2$ as the major gas flow in the growth chamber, we can control the ratio $P_{CH4}/P_{H2}$ to optimize the growth of BLG.

Figure 2 shows the representative SEM images for single crystalline MLG and BLG samples grown on the backside of Cu-foils, where the $P_{CH4}/P_{H2}$ ratios for MLG and BLG growth under the same $H_2$-flow rate (2.5 sccm) are 0.1 and 0.04, respectively. In the case of graphene samples grown on Cu-foils without extra processing except rinsing in IPA, the average lateral dimension of the single crystalline MLG flakes after 3 minutes of PECVD process is found to range from 2 μm to 5 μm, whereas the average diameters of the first- (light gray region) and second-layer (dark gray region) graphene in the BLG samples are around 3 μm and 1 μm, respectively, as shown in Figs. 2(a) and 2(b). In contrast, for Cu-foils pretreated by extra planarization processes (see Fig. 2 captions), the average domain sizes of the single crystalline MLG and BLG samples after 20 minutes of PECVD process are found to be around 15 μm and 5 μm, respectively, as shown in Figs. 2(c) and 2(d). Graphene growth under this condition was used for the construction of Raman spectroscopic database because the crystalline sizes thus obtained are sufficiently large to satisfy the resolution limit of our Raman spectrometer.

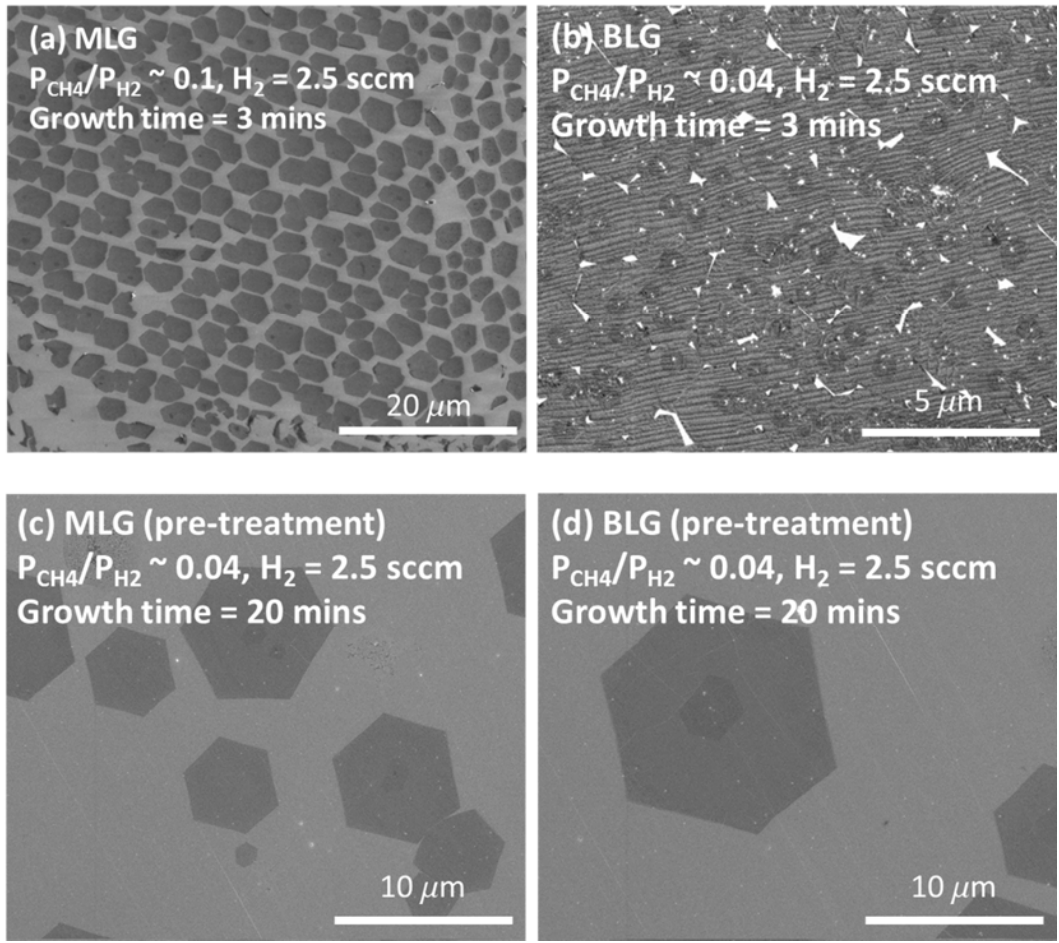

**Fig. 2.** SEM images for single crystalline MLG and BLG samples grown by PECVD on Cu foils. The lighter and darker gray colors represent the imaging contrasts between regions of MLG and BLG, respectively. **(a) -- (b)** Graphene grown on Cu foils that have been prepared by simple sonication in acetone and IPA for 5 minutes before insertion into the growth chamber. Single crystalline MLG or BLG were first grown on the Cu foil and then gradually coalesced into a large MLG or BLG films with increasing growth time. **(c) -- (d)** Graphene grown on pretreated Cu foils that have been first cleaned by acetic acid to remove $CuO_x$, sonicated in acetone and IPA for 5 minutes, and then inserted into furnace for annealing at 1050 °C in Ar and $H_2$ environment for 35 minutes to achieve planarization of the Cu surface, and finally placed into the PECVD system for graphene growth. The growth time was 20 minutes with $P_{CH4}/P_{H2} \sim 0.04$.

For Raman spectroscopic studies, graphene samples are separated from the Cu-substrates and transferred onto $SiO_2$/Si-substrates (Si wafers with a layer of 285 nm-thick thermal dry oxide on top) by means of the bubbling transfer method in buffered NaOH solution with the aid of PMMA [63], which ensures that our graphene samples are free of metal residues after the transfer. The Raman spectra of MLG and various BLG samples are exemplified in Figs. 3(a) and 3(b), where apparent variations in the Raman spectra are found in tBLG samples with different twist angles. The Raman modes for the G- and 2D-bands of MLG are located around 1584 cm$^{-1}$ and 2685 cm$^{-1}$, respectively, whereas those for the BLG samples are located around 1585 cm$^{-1}$ and 2700 cm$^{-1}$, respectively. The slight deviation of the G-band with respect to the typical value of 1582 cm$^{-1}$ is attributed to excess carrier doping from the environment (such as atmospheric adsorption of $H_2O$ and $O_2$) [36, 64] or the substrate [65]. Here we remark that the peak position and FWHM of each Raman mode have been obtained by fitting the peak to a single Lorentzian function. For MLG, the 2D-to-G intensity ratio ($I_{2D}/I_G$) and the FWHM of the G-, and 2D-bands are found to be ~ 4, 25 cm$^{-1}$ and 40 cm$^{-1}$, respectively. For BLG with AB stacking ($\theta = 0°$), the ($I_{2D}/I_G$) ratio and the FWHM of the G- and 2D-bands are ~ 0.6, 21 cm$^{-1}$ and 60 cm$^{-1}$, respectively. For tBLG with $\theta = 30°$, the ($I_{2D}/I_G$) ratio and the FWHM of the G- and 2D-bands are ~ 4, 17 cm$^{-1}$, and 30 cm$^{-1}$, respectively.

Within the tBLG samples, different interlayer twist angles are found and are manifested in the Raman spectroscopy by different ($I_{2D}/I_G$) ratios, different degrees of enhancement in the G-band intensity that are associated with the resonant electron-hole pairs between the π- and π$^*$ van Hove singularities (VHS) in the density of states, and the appearance of twist-angle dependent R- and R′-bands, where R- and R′-bands are the results of static interlayer potential-mediated inter- and intra-valley double-resonance Raman scattering processes [57, 66]. Here we note that the VHS arise from the overlap of two Dirac cones, and that optical excitation processes take place at the VHS will contribute to the G-band intensity [55,67,68]. Generally, the intensity of the G-band in BLG is slightly stronger than that in MLG due to the interferences of laser light in the stacked layers and multi-reflections of light when the number of graphene layers is smaller than 10 [69,70]. Similarly, different interlayer orientation changes the linewidth and position of the 2D-band; the 2D-band in the BLG is blue-shifted relative to that in MLG, and the blue-shift is non-monotonic with increasing twist angle.

Figure 3(a) shows the Raman spectra of MLG, AB-BLG ($\theta = 0°$), and tBLG with twist angle $\theta = 1.2°$, 4° and 30°. The FWHM of the 2D-band of AB-BLG and 30°-tBLG is found to be 60 cm$^{-1}$ and 25 cm$^{-1}$, respectively. The tiny D-band in the Raman spectrum of 30°-tBLG (left inset in Fig. 3(a)) is attributed to the interlayer coupling [56, 57] rather than from the present of defects, because the 2D-to-G intensity ratio of the 30°-tBLG sample (~ 4) is much larger than that of the MLG (~ 2), and the FWHM of the 2D-band for the 30°-tBLG sample is also slightly smaller than that of MLG (35 cm$^{-1}$). In contrast, the FWHM of the 2D-band for any tBLG sample with a twist angle smaller than 30° is generally broader.

Additionally, the absence of defective peaks such as the D- and D′-bands at around 1350 cm$^{-1}$ and 1620 cm$^{-1}$ for the tBLG sample with $\theta = 4°$ affirms the appearance of the R′-band at 1626 cm$^{-1}$ (right inset in Figure 3(a)), where the R′-band can only appear at a small twist angle (3° < $\theta$ < 9°) and becomes inaccessible at a large twist angle. This is because the latter would require a large reciprocal rotational vector $q$, where $q \equiv |\mathbf{b}_{1(2)} - \mathbf{b}'_{1(2)}|$ is the absolute value of the difference between the reciprocal lattice vectors $\mathbf{b}_{1(2)}$ and $\mathbf{b}'_{1(2)}$ for the top and bottom layers, respectively [41, 56, 66, 71, 72]. Although the D′-band associated with the disordered sp$^2$ carbon and the rotated angle dependent R′-band occur within the same Raman spectral range, we may exclude the possibility of D′ because its appearance must arise from high defect densities [73, 74].

Finally, Raman spectra for tBLG with the interlayer orientation around the resonant critical angles (12° and 14°) are plotted separately in Fig. 3(b), where the resonance due to VHS significantly enhances the intensity of the G-band so that the Raman spectra of other tBLG in Fig. 3(a) would have been difficult to resolve if they were plotted together on the same scale. The $\theta$-dependent, non-dispersive R-band located in the range of 1400 cm$^{-1}$ ~1500 cm$^{-1}$ is also visible for the Raman spectra near the resonant conditions, as marked in Fig. 3(b). All these Raman spectral characteristics are found to be consistent with previous reports [55, 59, 60].

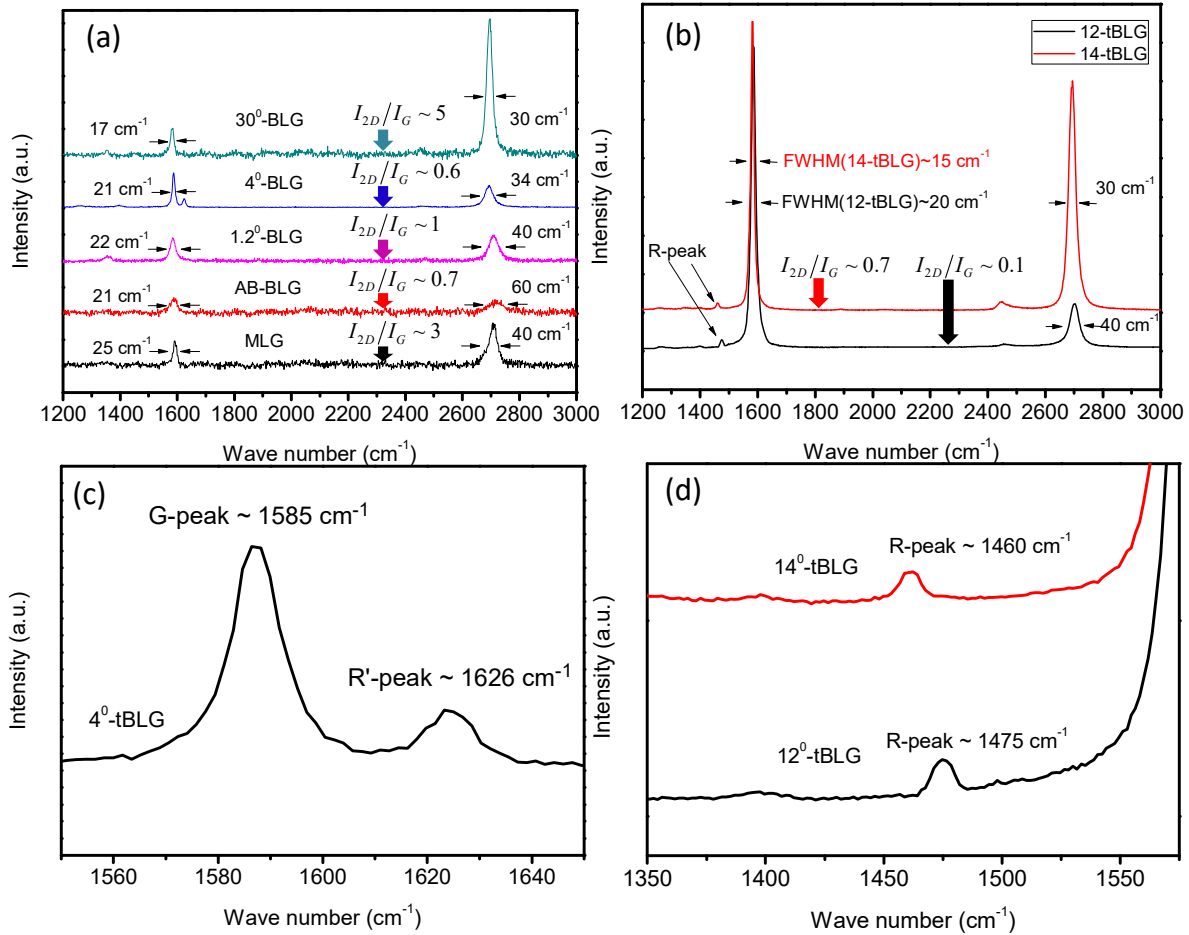

**Fig. 3. (a)** The Raman spectra for MLG, AB-BLG, and tBLG samples with $\theta = 1.2°$, $4°$ and $30°$. A non-dispersive D-band [57] is found in the Raman spectrum of $30°$-tBLG. **(b)** The Raman spectra for tBLG samples with $\theta = 12°$ and $14°$. **(c) -- (d)** Twist-angle dependent Raman spectral features of R- and R′-bands, where R- and R′-bands occur due to static interlayer potential-mediated inter- and intra-valley double-resonance Raman scattering processes. The R-band appears between 1400 cm$^{-1}$ and 1500 cm$^{-1}$ and R′-band positions around 1625 cm$^{-1}$.

In general, for a given excitation laser energy ($\sim 2.41$ eV), the Raman spectroscopy for BLG with an interlayer twist angle larger than the critical angle will exhibit characteristics resembling those of the Raman spectrum of MLG because all optical excitation processes effectively occur in an isolated Dirac cone. On the other hand, for tBLG with twist angles smaller than the critical angle, the resulting Raman spectra will be significantly different from those of MLG because the two closely spaced Dirac cones associated with the bilayers in the momentum space will give rise to significantly different scattering paths [55].

To ensure that the majority of the PECVD-grown tBLG samples are controlled within the small-angle configurations, the ($P_{CH4}/P_{H2}$) ratio during the PECVD growth must be kept to be sufficiently small, because small ($P_{CH4}/P_{H2}$) ratios can help reduce the growth rate of BLG and increase the interlayer coupling, which tends to favor the formation of smaller-angle tBLG (including the AB-BLG) [28, 58, 75-77]. After PECVD synthesis, the quality of the MLG and BLG samples is verified by noting the absence of the D-band and the presence of well-defined G-band and 2D-band characteristics in the Raman spectra. To determine the twist angle of the BLG single crystals, we directly measure the hexagonal crystalline edges between the first and second layers in the SEM images [56], and also analyze the peak positions of rotational R- and R′-bands in the Raman spectra [66].

In addition to direct growth of graphene on Cu-foils that are not pretreated except for IPA rinsing, we explore an alternative and more cumbersome way in an effort to enlarge the grain size of single crystalline BLG samples. Specifically, we subject the Cu-foils to two steps of pretreatment by first soaking them in acetic acid ($CH_3COOH$) to remove the surface $CuO_x$, and then annealing them in Ar and $H_2$ environment for 30 mins at 1050 °C to smooth the surface of the Cu-foils. These pretreated Cu-foils are subsequently inserted into the PECVD system for graphene growth. (See Supplementary Note 2 for details of the pretreatment procedure). Using these pretreated Cu-foils as the growth substrate, the average grain sizes of PECVD-grown MLG and BLG after 20-min become 15 μm and 5 μm, respectively. The edges of these larger samples become clearly visible under an optical microscope, which makes the identification of the twist angle much easier. Additionally, we note that BLG samples grown on these pretreated Cu-foils do not exhibit a preference for small twist angles.

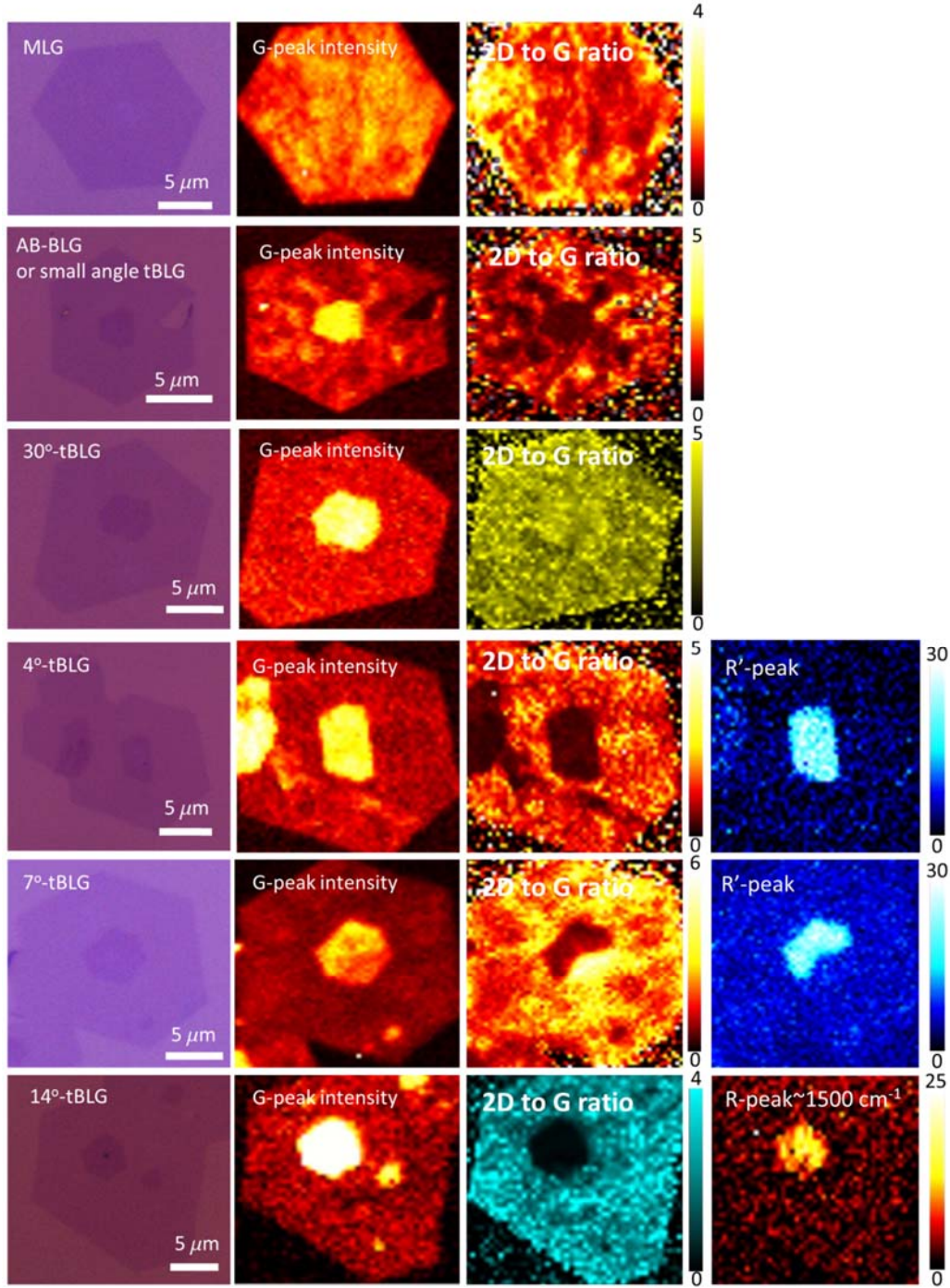

**Fig. 4.** Optical images and spatially resolved Raman maps taken on MLG and BLG samples with varying twist angles. The Raman maps include maps of the G-band intensity, 2D-to-G intensity ratio, and the R- and R′-bands whenever they appear for specific twist angles. The peculiar R′-band mapping shown in the case of 7°-tBLG indicates that only one half of the BLG has $\theta = 7°$ whereas the other half of the BLG has $\theta = 30°$, as confirmed by the point Raman spectra. In the case of 14°-tBLG Raman mapping, the R-band indicates inter-valley scattering, which only appears in the BLG region.

Using these larger BLG single crystals, we proceed to construct a database for the Raman spectra of tBLG. Figure 4 shows a collection of the optical images and spatially resolved Raman spectroscopic maps for samples of MLG, very small-angle tBLG (with $\theta < \sim 1°$), 4°-tBLG, 7°-tBLG, 14°-tBLG and 30°-tBLG.

*2.2.2 Growth of Large Bilayer Graphene Films*

To achieve the growth of large BLG films by PECVD, we follow the same process as that described in Subsection 2.2.1 for the growth of tBLG single crystals with small twist-angle distributions on Cu-foils that have only been sonicated in acetone and IPA, except that the $H_2$ flow rate is increased from 2 sccm to 3 sccm, and the growth time is increased from 3 minutes to 10 minutes.

Figure 5(a) – 5(d) show the optical microscope (OM) images for large MLG and BLG films transferred onto the 285 nm-thick $SiO_2$/Si substrates by means of a polymer-free transfer method [78]. Specifically, Fig. 5(a) shows the OM of a large MLG film synthesized under an $H_2$-flow rate of 2 sccm and a gas ratio ($P_{CH4}/P_{H2}$) = 0.1. Under such conditions, the sample is predominantly MLG although occasionally small areas of FLG (on the order of 100's nm$^2$) can be found. The predominant MLG films are always continuous and cover the Cu foil fully without apparent grain boundaries, which is consistent with our previous findings [17]. Figures 5(b) – 5(d) as well as Fig. S3 exhibit the evolution of the BLG film growth with different ratios of ($P_{CH4}/P_{H2}$) over the same growth time of 20 minutes. (More details about the growth conditions are given in Supplementary Note 2.) The slight color contrasts between MLG and BLG in Figs. 5(b) and 5(c) suggest that the density of the second layer (grown underneath the first layer) and the grain size increases with the increase of ($P_{CH4}/P_{H2}$), and Fig. 5(d) reveals a largely uniform BLG film for ($P_{CH4}/P_{H2}$) = 0.06. With further increase of the ($P_{CH4}/P_{H2}$) ratio (> 0.06), however, the films exhibit non-uniform thicknesses over the whole Cu-foil, with a typical domain size around few tens of μm$^2$. On the other hand, increasing the $H_2$-flow rate (> 5 sccm) always results in large MLG films.

These findings suggest that the growth of large and continuous BLG films can only be achieved within a very small parameter space.

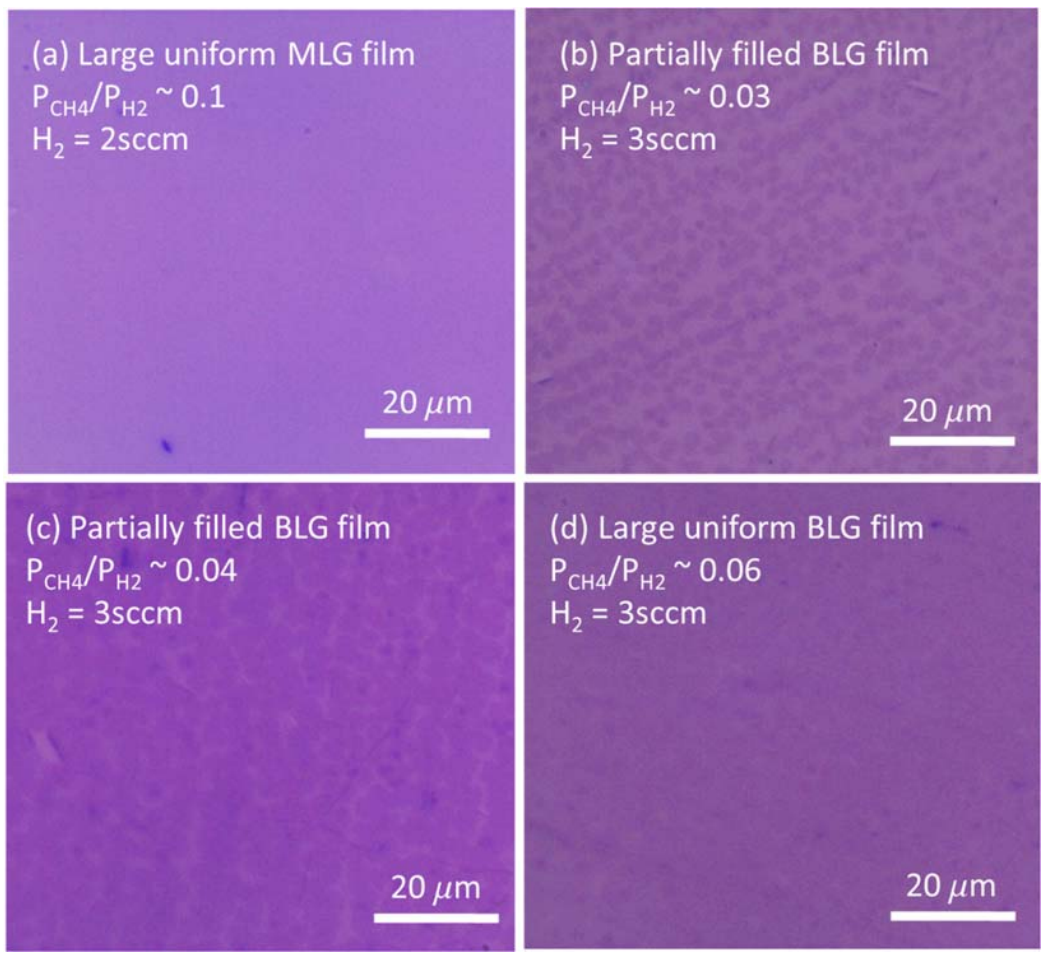

**Fig. 5.** Optical images of large MLG and BLG films transferred onto 285 nm-thick SiO$_2$/Si substrates and grown under different ($P_{CH4}$/$P_{H2}$) ratios over the same growth time: **(a)** a large MLG film, **(b)** a partially filled BLG film with smaller coverage under a large MLG, **(c)** a partially filled BLG film with higher coverage under a large MLG, and **(d)** a large uniform BLG film. Additional SEM images for the evolution of BLG growth with different PECVD growth conditions can be found in Fig. S3, and detailed SEM images and Raman point spectra taken on an mm-size uniform BLG film are given in Fig. 6 and Fig. S4.

In Figure S5 and Table S1 we summarize how the growth of tBLG and the twist angle between layers of the BLG evolve with the partial pressure ratio ($P_{CH4}$/$P_{H2}$). Additionally, we note that the H$_2$ flow rate used for sizable BLG growth in our PECVD system must be limited to a small range from 2.5 sccm (corresponding to P$_{H2}$ ~ 1.6×10$^{-5}$ torr) to 3 sccm (corresponding to P$_{H2}$ ~ 2×10$^{-5}$ torr). Within this limited range of H$_2$ flow rate, a larger H$_2$ flow rate would lead to a slightly larger partial pressure ratio (P$_{CH4}$/P$_{H2}$) necessary to control the twist angle but would not significantly alter the trend of the control of the twist angle, as detailed in Supplementary Note 2, Figure S5 and Table S1.

Figure 6(a) is an optical image of an mm-size BLG film (grown with H$_2$ flow rate = 3 sccm and the partial pressure ratio (P$_{CH4}$/P$_{H2}$) = 0.06) transferred onto and fully-covered a 285nm-thick SiO$_2$/Si substrate. Figures 6(b) and 6(c) are optical micrographs of selected regions in Fig. 6(a) with 5× and 100× magnifications, respectively. Raman spectroscopic characterization of this mm-size BLG film has been conducted on more than 50 points at random locations, and most spectra reveal similar Raman spectral characteristics that are consistent with small twist angles, as exemplified by 10 different point spectra in Fig. 6(d). To verify the reproducibility of the growth conditions, we have synthesized more than 50 samples with the same growth parameters and found that the growth of mm-sized BLG films with small twist angles configuration is indeed reproducible, as further exemplified in Fig. S4. By comparing the Raman spectra taken on this mm-size BLG film in Fig. 6 and Fig. S4 with those shown in Fig. 4, we find that the Raman spectral characteristics are mainly consistent with the small-angle configuration, which may be either AB-BLG or a very small angle tBLG with $\theta <\sim 1°$.

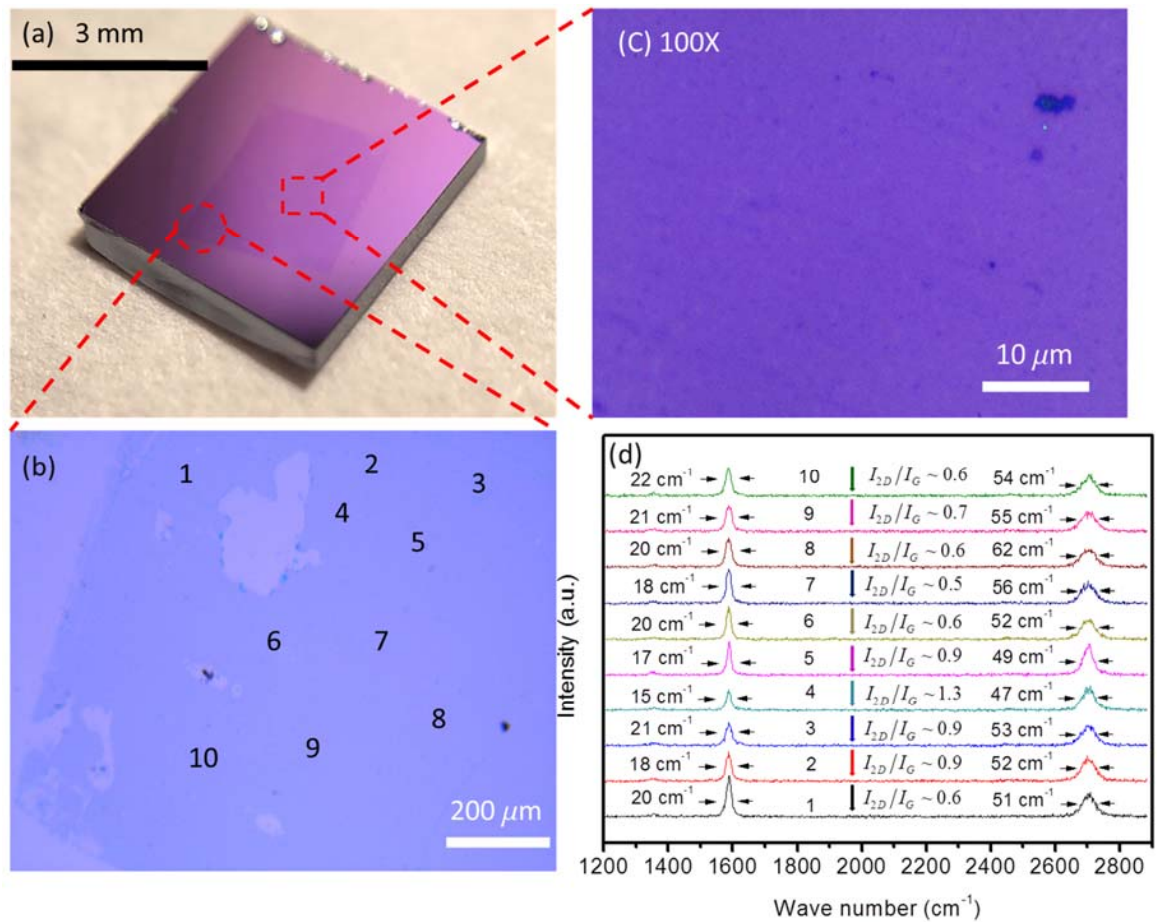

**Fig. 6. (a)** Image of a large BLG film transferred onto a 285 nm-thick SiO$_2$/Si substrate by means of a polymer-free transfer method. **(b)** and **(c)** are OM images for selected regions of large BLG film with magnification (scale bar size) 5× (200 μm) and 100× (10 μm), respectively. **(d)** Raman spectra taken at different ten points labeled in (b), showing consistent spectral characteristics.

To further verify the consistency of sample quality throughout large-area MLG and BLG films, we conduct spatially resolved Raman mapping on these samples. The Raman mapping data are collected using Reinshaw @ Invia 514 nm with laser power intensity at 2 mW, data acquisition time in 1 s, a diffraction-limited spot size of ~ 500 nm, and the pixel step of 0.5 μm over an area (50 × 50) μm$^2$. The FWHM of the 2D-band is extracted by fitting the Raman peak to a single Lorentzian function, and the resulting FWHM is found to be in the range of 10 cm$^{-1}$ to 40 cm$^{-1}$ (30 cm$^{-1}$ to 60 cm$^{-1}$) for MLG (BLG) films. Figure 7 shows representative

Raman maps of the G-band intensity, 2D-band linewidth, and the 2D-to-G intensity ratio of both large-area MLG and BLG films. In general, the average G-band intensity in BLG films is larger than that in MLG films. The overall broadening of 2D linewidth (> 40 cm$^{-1}$) and smaller 2D-to-G intensity ratio (< 1.2) in the BLG films suggest that most BLG areas are consistent with the small-angle configuration.

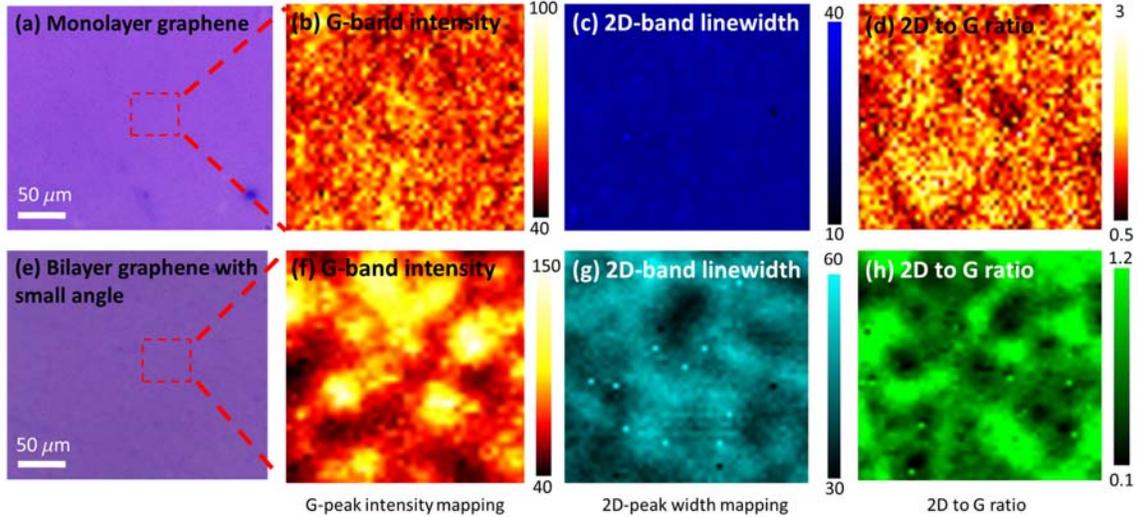

**Fig. 7**. Optical images and spatially resolved Raman maps of large MLG and BLG films: **(a)** Optical image of a large MLG film. **(b)** – **(d)** Maps of the G-band intensity, 2D-band linewidth and 2D-to-G intensity ratio for the MLG area indicated by the red dashed box in (a). **(e)** Optical image of a large BLG film. **(f)** – **(h)** Maps of the G-band intensity, 2D-band linewidth and 2D-to-G intensity ratio for the BLG area indicated by the red dashed box in (e).

Although Raman spectroscopy is a powerful and nondestructive tool for identifying the basic properties of graphene systems (including MLG, AB-BLG, tBLG, and FLG), it still lacks sufficient spatial resolution to pinpoint the exact interlayer twist angle of a tBLG sample. To address the issue of spatial resolution, we conduct atomically resolved scanning tunneling microscopic studies on some of the BLG samples. As exemplified in Fig. S6, we obtain spatial images of the Moiré patterns on two tBLG samples and determine the twist
21

angles ($\theta = 13°$ and $\theta < 1°$) from the Moiré period $\lambda$ by the relation $\cos\theta = 1 - (a^2/2\lambda^2)$ [79], where $a$ denotes the graphene lattice constant. In the small twist angle limit, we find a simple relation $\theta \approx (a/\lambda)$ so that the Moiré period increases with decreasing twist angle [79].

## 3. Measurement of the Work Function by KPFM and UPS

Kelvin probe force microscopy (KPFM) is a technique that enables nanoscale spatially resolved measurements of the surface potential difference between an atomic force microscopy (AFM) tip and the sample [80-82]. The contact potential differences (CPD) measured by the strength of the electrostatic forces between a conductive probe and the sample in the KPFM can reflect on the surface potential difference between two materials, which can be directly transformed into the work function [83].

We perform surface potential measurements of the PECVD-grown BLG samples by the KPFM based on the Bruker Dimension Icon AFM system (Model: PFQNE-AL). To prevent possible charge transfer between the substrate and graphene due to extrinsic effects such as electron-hole puddles [84], doping domains in graphene [85], and quantum capacitances [86], we first transfer the PECVD-grown single crystalline BLG flakes onto the Au (111)/mica substrates. Measurements of the BLG surface potential are then carried out on the BLG/Au(111)/mica sample by KPFM in the ambient environment. Throughout the KPFM measurement, the distance between the AFM tip and sample has been fixed at 5 nm and the Au (111)/mica substrate grounded. The surface potential difference thus determined can be converted to the work function ($\varphi$) of BLG because the work function of Au (111)/mica has been calibrated to be 4.7 eV by ultraviolet photoelectron spectroscopy (UPS). Thus, the work function of BLG is given by $|\varphi| = |(4.7 \text{ eV}) - (\text{surface potential difference})|$ [82].

The main panel of Fig. 8(a) shows a spatially resolved surface potential map of a sample obtained by the KPFM under ambient environment and with the pixel steps at 0.5 μm, and the inset is a Raman map of the G-band intensity on the same sample. In Fig. 8(b), representative Raman spectra are shown for the MLG region (green cross) and the nearly AB-stacking BLG region (solid-blue circle). The absence of any discernible D-band intensity suggests the high quality of our PECVD-grown graphene single crystals. The work function measured along the red line in Fig. 8(a) is illustrated in Fig. 8(c), where the work functions of MLG and BLG are found to be 4.48 eV and 4.58 eV, respectively. Both values are consistent with results reported previously [87-91]. We note that the surface potential difference between MLG and BLG is around 100 mV, which is comparable to previous reports of the graphene surface potential increasing with the number of layers [45, 87, 92, 93]. Importantly, we do not find any discernible surface potential differences for tBLG with different interlayer twist angles.

We have also performed UPS measurements on large MLG and BLG films and found that their work functions are 4.65 eV and 4.72 eV [94-96], respectively, as shown in Fig. 8(d). The work function measurements for MLG and BLG large films are performed via normal emission UPS by using monochromatic He-I radiation with 21.2 eV photon energy as the UV source. The UPS characterization has been carried out in ultra-high vacuum (UHV), and the work function difference between MLG and BLG is found to be ~ 70 meV, which is on the same order of magnitude as the results obtained from KPFM. The slight difference between the results from KPFM and UPS may be attributed to environmental issues, such as atmospheric humidity for KPFM studies in the ambient environment.

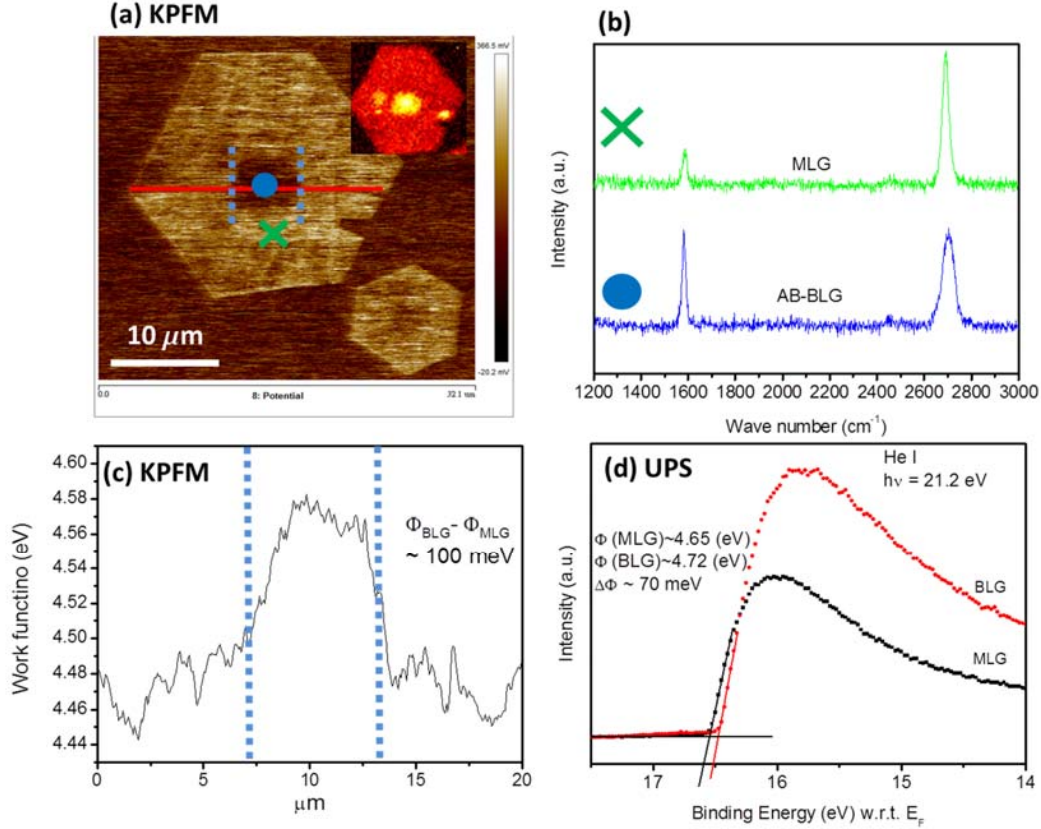

**Fig. 8. (a)** Mapping of the surface potential of a large graphene film by KPFM. The inset shows the corresponding Raman map of the G-band intensity of the same sample. **(b)** Raman point spectra taken on an MLG area (green cross) and a nearly AB-stacking BLG area (blue solid-circle). **(c)** Spatial evolution of the work function measured along the red line depicted in (a), showing an apparent increase of the work function in the BLG area. **(d)** UPS measurements of the work functions of both MLG and BLG films, yielding values of 4.65 eV and 4.72 eV, respectively.

## 4. Electrical Measurement

In this subsection, we describe our investigation of the back-gated two-terminal field-effect transistor (FET) devices based on the PECVD-grown BLG samples in order to elucidate their electronic transport properties.

PECVD-grown single crystalline BLG flakes and large BLG films are transferred onto (100) $p^{++}$-Si chips (with resistivity ranging from 0.001 to 0.005 Ω-cm), which serve as the gate electrode with a capping layer of 285 nm-thick $SiO_2$. The carrier density ($n$) in the BLG can be tuned by a back-gate voltage ($V_g$) via the relation $n = C_{ox}/e(V_g - V_{CNP})$, where $C_{ox} = 11.6$ nF/cm$^2$ is defined as the capacitance per unit area of the 285 nm-thick SiO$_2$, and $V_{CNP}$ indicates the shift of charge neutrality point ($V_{CNP}$) originated from unintentional doping in BLG. Nearly AB-stacking BLG single crystalline flakes are chosen as the material for the two-terminal back-gated device fabrication. For large BLG films, we also intentionally selected BLG region with small twist angles for device fabrication.

To fabricate the two-terminal devices, source and drain contacts are defined by electron-beam lithography. A combination of Pd (10 nm thick) and Au (40 nm thick) are thermally evaporated onto the source and drain as contact electrodes. The BLG channel with a 1 μm-width ($W$) and a 3 μm-length ($L$) is subsequently defined by electron-beam lithography and then etched by reactive ion etchant (RIE). The optical micrograph of a representative two-terminal device is shown in Fig. 9(a), where the two black lines indicate the BLG channel with the aforementioned dimensions, and the inset shows schematics of the device cross-section. Prior to the electrical measurement, lithographically patterned devices are baked in the forming gas environment (Ar and H$_2$) at 350 °C to remove chemical residues (such as PMMA residues remained on the BLG channel during the sample transfer and electron-beam lithography process) and atmospheric adsorbents like H$_2$O and O$_2$. Electrical measurements on the two-terminal devices are taken at room temperature and in a vacuum environment (@ ~ 20 mTorr).

Figures 9(b) and 9(c) show representative measurements of the source-drain conductivity ($\sigma$) and resistivity ($\rho$) of BLG samples as a function of the gate voltage ($V_g$) in the vicinity of the charge neutrality point (CNP), where the black (red) curve represents the data taken on a back-gated FET device based on a single crystalline AB-BLG flake (a large BLG film). The source-drain voltage ($V_{sd}$) has been kept at 10 mV for the electrical measurements, which ensures that the corresponding back-gate electric field ($E_{bg}$) is restricted within the range of ±2.3 MV/cm to prevent from possible breakdown of the SiO$_2$ dielectric layer. The back-gate voltage ($V_g$) is swept from −70 V to 70 V, with the drain current ($I_{sd}$) passing through the BLG channel measured simultaneously. Thus, the conductivity ($\sigma$) can be obtained via the relation $\sigma = \frac{L}{W}\frac{I_{sd}}{V_{sd}}$. We have examined over 20 BLG devices and all of them exhibit the V-shaped conductance ($\sigma$) as a function of $V_g$.

The ambipolar behavior revealed in the electrical transport measurements in Fig. 9(b), where the conductivity in the negative back-gate voltage regions exhibits higher values relative the positive back-gate voltage regions, may be attributed to the asymmetric effects of Fermi-energy sweep across p-p and p-n junctions in the BLG. Representative current on-off ratios, *i.e.*, ($I_{on}/I_{off}$), for devices based on the single crystalline BLG flake and the large BLG film are found to reach around 8 and 5, respectively. Additionally, the shift of charge neutrality point (CNP) is found to be around 25 V for the device based on the single crystalline BLG flake and around 28 V for the device based on the large BLG film. The shift in CNP could involve influence from substrate-induced electron-hole puddles, ripples, and residual atmospheric adsorbates [97].

The slight difference in conductivity (and also resistivity) between the single crystalline BLG flake and the large BLG film may be attributed to the presence of structural disorder, higher concentrations of nucleation centers, or even wrinkles caused short-range scattering encountered in the large BLG film-based devices. The sub-linear behavior at high values of $V_g$, which corresponded to saturation of conductivity in the limit of higher carrier density, as shown in Fig. 9(b) for both groups of BLG samples, may be attributed to the presence of short-range scattering sites (such as atomic defects) [98-102]. In contrast, long-range scattering mechanisms associated with charged impurity scattering and electron-hole puddles are known to dominate at low carrier density (*i.e.*, at low $V_g$) and determine the minimum conductivity at CNP [98-102].

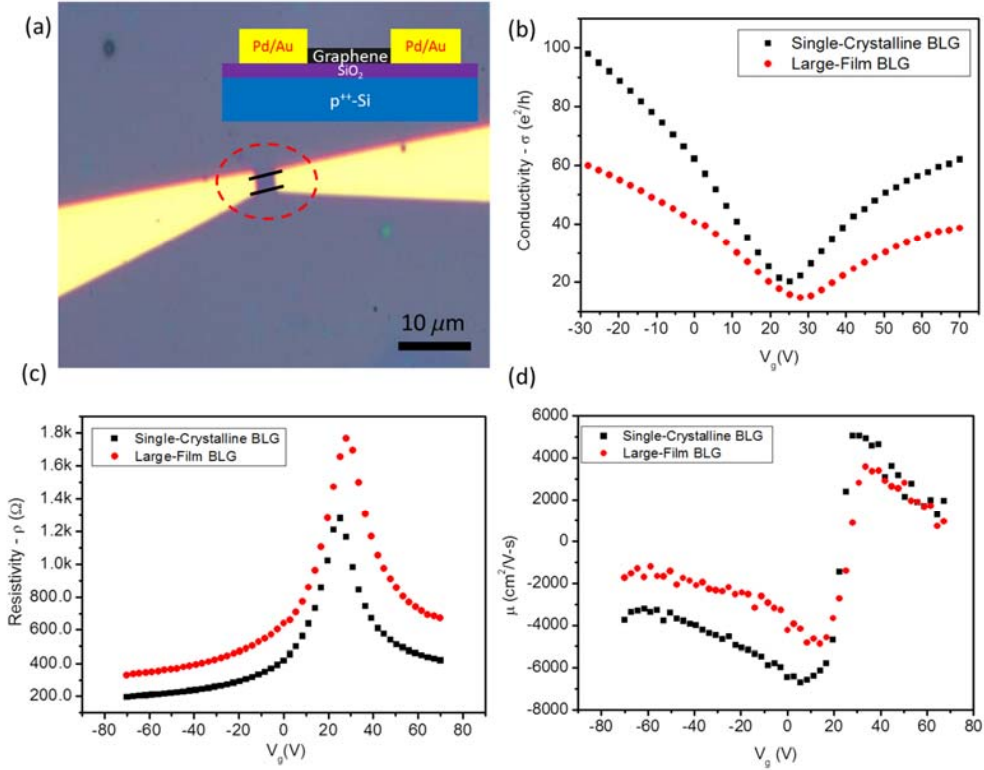

**Fig. 9.** Electrical transport measurements of BLG samples: **(a)** <u>Main panel</u>: optical micrograph of a two-terminal back-gated FET device based on a BLG sample. The area enclosed by two black lines represents the BLG channel with a 3-μm length and a 1-μm width. <u>Inset</u>: Schematic side-view of the FET device. **(b)** Representative conductivity versus gate-voltage (σ-vs.-$V_g$) curves for a single crystalline BLG (black squares) and a large BLG film (red circles), measured in units of ($e^2/h$). **(c)** Representative resistivity versus gate-voltage (ρ-vs.-$V_g$) curves for a single crystalline BLG (black squares) and a large BLG film (red circles). Both σ-vs.-$V_g$ and ρ-vs.-$V_g$ data have been taken with the source-drain voltage ($V_{sd}$) kept constant at 10 mV. **(d)** Carrier mobility of both the single-crystalline BLG (black squares) and large BLG film (red circles) derived from σ-vs.-$V_g$ measurements and the direct-transconductance method. For single crystalline AB-BLG, the highest electron and hole mobility values are 5000 cm$^2$/Vs and 7000 cm$^2$/V-s, respectively. For large-film BLG, the highest electron and hole mobility values are 4000 cm$^2$/V-s and 5000 cm$^2$/V-s, respectively.

The carrier mobility ($\mu$) of the BLG is also extracted by employing the direct-transconductance method via the relation $\mu \approx \dfrac{1}{C_{ox}} \dfrac{d\sigma(V_g)}{dV_g}$, where the mobility is measured in units of cm$^2$/V-s, as shown in Fig. 9(d). The carrier mobility in the vicinity of CNP for both groups of BLG devices is found to be in the range of 4000 to 7000 cm$^2$/V-s, which is consistent with previous reports for BLG on SiO$_2$/Si [13, 103], although these values are generally smaller than those found in MLG [17, 104-106].

Figure 10 shows the temperature dependent electrical measurements of a two-terminal FET device based on a nearly AB-stacking BLG single crystal from 300 K to 100 K. It is known that the temperature dependence of the conductivity behavior of graphene systems (both MLG and BLG) consists of contributions from several competing factors, including thermal activation of carriers from valence band to conduction band, electron-phonon scattering, temperature dependent screening effects, and activated carriers that are locally and thermally excited above the random-distributed charged disorder [104, 107, 108]. For a perfect AB-stacking BLG known as a zero-gap semiconductor, direct thermal excitations of carriers from valence band to conduction band can comprise of an important contribution to the temperature dependent transport, which lead to reduced conductivity with decreasing temperature in the vicinity of CNP [108-112] analogous to the findings in MLG [108, 113, 114]. On the other hand, the insulating temperature dependence of the conductivity at high carrier densities in our BLG samples suggests that the dominant transport mechanism is thermally mediated, which differs from the metallic conductivity of MLG at high carrier densities [108, 113, 114]. Finally, we note that for very small twist angles near the magic angle, flat Moiré bands with a small energy gap emerge [27] so that an insulator-to-superconductor transition can be induced by tuning the gate voltage at very low temperatures [28]. Further systematic studies of the electrical transport properties of our PECVD-grown small-angle tBLG samples at low temperatures as a function of the twist angle will not only help verify the quality of these samples for superconductivity but also elucidate the evolution of transport mechanisms with changing Moiré bands.

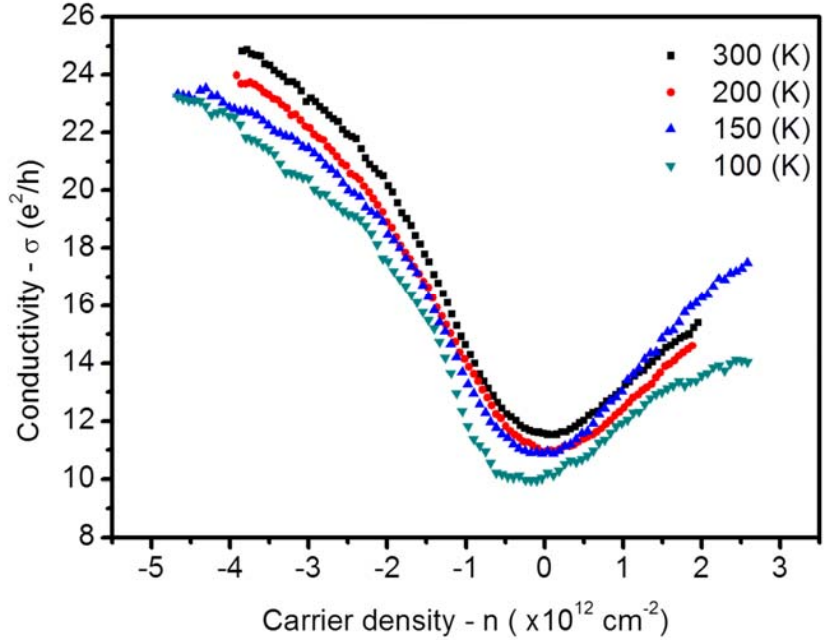

**Fig. 10.** Conductivity σ of a back-gated FET device based on a nearly AB-stacking BLG single crystal is measured as a function of the carrier density $n = C_{ox}/e(V_g - V_{CNP})$ at various constant temperatures from 100 K to 300 K, where $C_{ox}$ = 11.6 nF/cm$^2$. Insulating temperature dependence is found at both low and high carrier densities for this BLG device. Here we have horizontally shifted each isotherm of $\sigma(n)$ curve to the charge neutrality point (CNP) where $n = 0$ for easier comparison of the conductivity at different temperatures.

**5. Conclusion and Outlook**.

In conclusion, we report in this work our development of a direct PECVD method to achieve efficient growth of single crystalline BLG hexagons and large BLG films (mm-size) with small twist angles by precise control of the growth parameters, particularly the CH$_4$-to-H$_2$ pressure ratio, (P$_{CH4}$/P$_{H2}$). Raman spectroscopic studies of these BLG samples reveal interesting twist-angle dependent spectral characteristics and suggest high sample quality. KPFM and UPS measurements of both types of BLG samples reveal a work function value that is consistent with the expected work function for pristine BLG and is also independent of the twist angle. Atomically resolved STM studies are conducted on the BLG samples to identify the Moiré pattern and the value of the corresponding twist-angle between the layers. Electrical transport measurements of more than 20 back-gated FET devices based on lithographically patterned BLG samples further demonstrate device characteristics that are consistent with high-quality BLG. Therefore, these results suggest a novel and efficient approach to investigate the physics and devices based on the Moiré bands of tBLG by controlled PECVD growth.

**Acknowledgments**

The work at Caltech was jointly supported by the National Science Foundation under the Institute for Quantum Information and Matter (IQIM) and the Army Research Office under the Multi-University Research Initiative (MURI) program. Y.-C. Chen, C.-D. Chen and Y.-S. Wu gratefully acknowledge the Dragon-Gate Program (MoST 106-2911-I-007-520) under the Ministry of Science and Technology (MoST) in Taiwan (R.O.C.) for supporting their visit to Caltech and the collaborative research. We thank Prof. Io-Chun Hoi, Dr. Ping-Yi Wen, Dr. Po-Hsun Ho, Dr. Chin-Pin Lee and master student Yen-Tsia Wang for their assistance with the low-temperature electrical transport measurements and the wire-bonding techniques; Prof. Mei-Ying Chou for her support during the execution of Dragon-Gate Project; Prof. Po-Wen Chiu, Dr. Chao-Hui Yeh, and PhD student Zheng-Yong Liang for their experimental guidance of Y.-C. Chen in graphene growth and fabrication of devices based on 2D materials; and Caltech PhD student Chen-Chih Hsu for useful discussion about PECVD growth techniques.

**Author contributions**

First author Y.-C. Chen investigated the entire PECVD growth procedures for the synthesis of BLG, carried out Raman spectroscopic characterization, SEM surface characterization, two-terminal back-gated devices fabrication, and related electrical measurements. W.-H Lin carried out KPFM measurements and also had in-depth discussions with Y.-C. Chen about PECVD-growth of MLG and BLG. W.-S. Tseng performed the work function measurements using UPS and also mentored Y.-C. Chen in the operation of the PECVD system. C.-C. Chen carried out STM measurements and related analyses. G. R. Rossman provided the Raman spectroscopic measurement facilities that are critically important for this work. C.-D. Chen and Y.-S. Wu engaged in active discussions about the experimental and theoretical development of this project. N.-C. Yeh as the principal investigator at Caltech coordinated the research project, data analysis, and wrote the manuscript together with Y.-C. Chen.

**Additional Information**

Supplementary Information:

    Supplementary Notes 1

    Supplementary Notes 2

    Supplementary Figure 1

    Supplementary Figure 2

    Supplementary Figure 3

    Supplementary Figure 4

    Supplementary Figure 5

    Supplementary Figure 6

    Supplementary Table S1

# Supplementary Information

## Direct growth of mm-size twisted bilayer graphene by plasma-enhanced chemical vapor deposition

Yen-Chun Chen,[a,b] Wei-Hsiang Lin,[c] Wei-Shiuan Tseng,[a] Chien-Chang Chen,[a] George. R. Rossman,[d] Chii-Dong Chen,[e] Yu-Shu Wu,[b,f] and Nai-Chang Yeh[a,g,*]

[a] Department of Physics, California Institute of Technology (Caltech), Pasadena, CA, 91125, USA

[b] Department of Physics, National Tsing-Hua University, Hsin-Chu 30013, Taiwan, ROC

[c] Department of Applied Physics, California Institute of Technology (Caltech), Pasadena, CA, 91125, USA

[d] Division of Geological and Planetary Science, California Institute of Technology (Caltech), Pasadena, CA, 91125, USA

[e] Institute of Physics, Academia Sinica, Nankang, Taipei 11529, Taiwan, ROC

[f] Department of Electronic Engineer, National Tsing-Hua University, Hsin-Chu 30013, Taiwan, ROC

[g] Kavli Nanoscience Institute, California Institute of Technology (Caltech), Pasadena, CA, 91125, USA

* Corresponding author. Tel: 626 395-4313. E-mail: ncyeh@caltech.edu (N.-C. Yeh)

Supplementary Notes 1
Supplementary Notes 2
Supplementary Figure 1
Supplementary Figure 2
Supplementary Figure 3
Supplementary Figure 4
Supplementary Figure 5
Supplementary Figure 6
Supplementary Table 1

# Supplementary Note 1: Experimental Setup

The PECVD system consists of a microwave plasma source, a growth chamber and a gas delivery system. The plasma source (Opthos Instruments Inc.) includes an Evenson cavity and a power supply (MPG-4) that provides an excitation frequency of 2450 MHz. The Evenson cavity matches the size of the growth chamber, which primarily consists of a 1/2-inch quartz tube (with the inner and outer diameters being 10.0 mm and 12.5 mm, respectively) and components for vacuum control. The reactant gas delivery system consists of four mass flow controllers (MFCs) for $CH_4$, Ar, $H_2$ and $O_2$. An extra variable leak valve is placed before the $CH_4$-MFC for precise control of the partial pressure of $CH_4$. During the growth process, the pressure of the system is maintained at ~ 25 mTorr. For the PECVD growth substrates, we use 25 μm-thick Cu-foils (Alfa Aesar with purity = 99.9996%). Prior to the graphene synthesis, the Cu-foils are always sonicated in ACE and IPA for 5 minutes then dried by nitrogen gas before inserted into the growth chamber. Several pieces of (1.5 × 0.8) $cm^2$ Cu-foils may be first placed on a quartz boat and then introduced into the growth chamber.

Figure S1 shows our experimental apparatus of the direct PECVD growth system without additional furnace. A (1.5 × 0.8) $cm^2$ Cu-foil was first placed on the quartz boat and introduced into 1/2-inch quartz made process tube (top panel), the plasma was subsequently ignited in a mixture of $CH_4$ and $H_2$ gas and graphene growth on Cu ensued (middle panel). The partial pressure of each gas was detected by a residual gas analyzer (RGA). After graphene growth (bottom panel), the sample was cooled to room temperature within at least 30 mins without breaking vacuum. Cu deposit on the quartz tube and holder after the graphene growth (bottom panel) was the result of Cu etching and was taken to be a signature for a successful run.

**Supplementary Figure 1:**

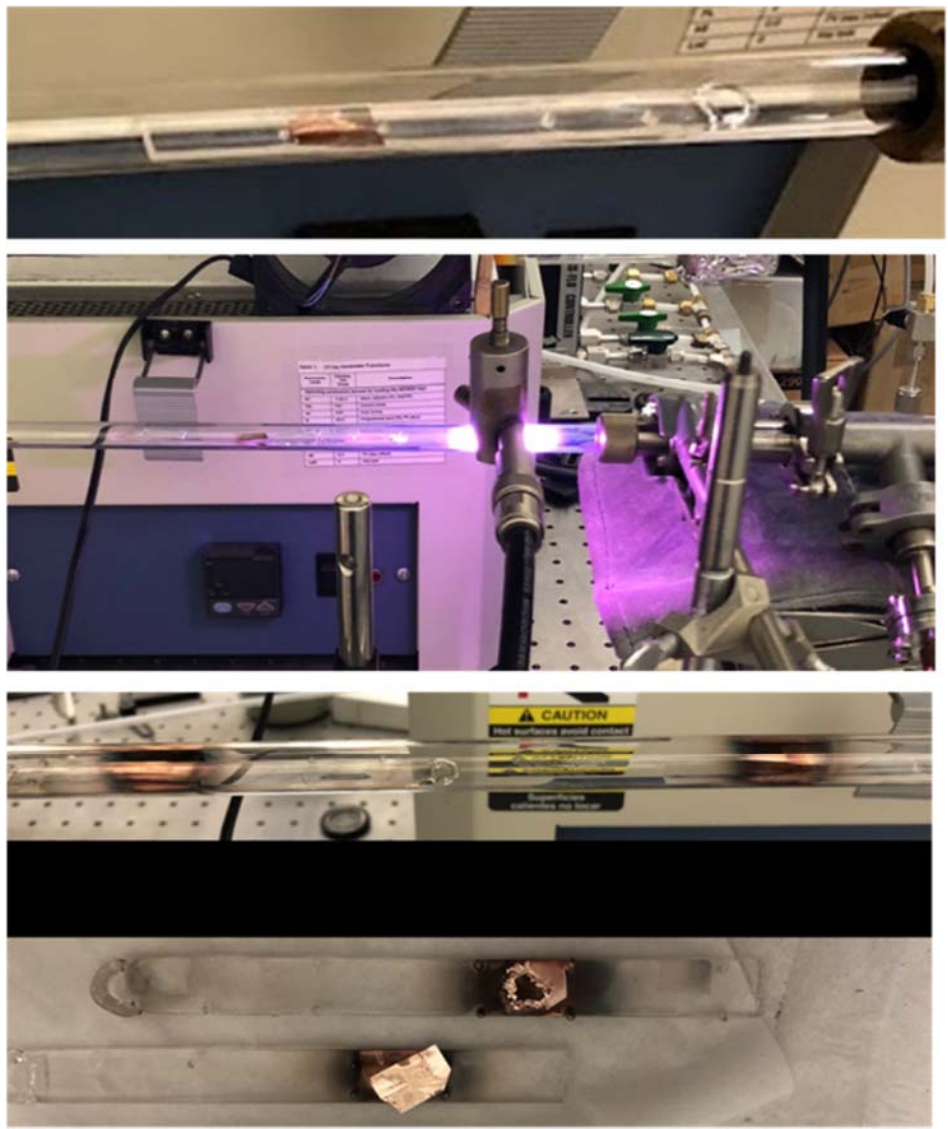

**Fig. S1.** Schematic of the PECVD setup. <u>Top panel</u>: Photograph showing a 25 μm thick polycrystalline Cu-foil (purchased from Alfa Aesar with purity 99.9996%) being placed on a quart boat and inserted into the PECVD growth chamber that consists of a ½-inch quartz tube and vacuum components. <u>Middle panel</u>: Photograph showing continuous exposure of a Cu-foil to microwave plasma in a mixture of $CH_4$ and $H_2$ gas for ensuing graphene growth on Cu. <u>Bottom panel</u>: Photograph showing Cu deposit on both quartz tube and quartz boat after graphene growth due to the etching of Cu-foil during the growth process. The deposition of Cu is taken to be a signature for a successful run.

# Supplementary Note 2: Sample Growth Conditions

This section describes the growth parameters for three types of BLG samples, and is divided into three parts: 1) growth parameters for hexagonal single crystalline BLG flakes, 2) sample preparation and growth parameters for large single crystalline BLG flakes, and 3) growth parameters for large BLG films with small twist-angle distributions.

The growth system is shown in Fig. S1. During the PECVD-growth of graphene, the system pressure is always kept at ~ $2\times10^{-2}$ torr, and the growth condition is mainly sensitive to the partial pressure of each reactant gas such as $CH_4$, $H_2$, and $N_2$ ($N_2$ is not controllable), but is insensitive to the background residual gases like Ar, $N_2$, $O_2$, and $H_2O$.

After inserting Cu-foils into the process tube, the system is first purged with Ar gas for at least 30 mins to ensure that the partial pressure of each residual gas remains largely the same. After the purge with Ar gas, $H_2$ is introduced into the system and then microwave plasma is ignited and kept on for 2 mins to remove $CuO_x$ on the Cu-foils under a controlled gas pressure of 500 mTorr. The reactant gases for graphene growth ($CH_4$ and $H_2$) are subsequently added into the PECVD system and controlled to a stable flow for 10 mins before the plasma is ignited and set to a power of 40 W for ensuing graphene growth. After the graphene growth, the sample is cooled to room temperature for at least 30 mins without breaking the vacuum.

## 1) Growth conditions for hexagonal single crystalline BLG flakes

Under the growth parameters provided below in Table 1, the typical grain size (~ the diameter of a hexagon) of purely single crystalline MLG is around 3 μm. For BLG, the averaged grain size for the first and the second-layer is 5 μm and 1 μm, respectively. These growth parameters for single crystalline MLG and BLG samples are summarized in Table 1:

| **Table 1** | Plasma (W) | $P_{CH_4}$ (torr) | $P_{H_2}$ (torr) | $P_{CH_4}/P_{H_2}$ | Growth Time |
|---|---|---|---|---|---|
| MLG | 40 | $1\times10^{-6}$ | $1.2\times10^{-5}$ (~ 2 sccm) | 0.1 | 3 mins |
| BLG | 40 | $7\times10^{-7}$ | $1.6\times10^{-5}$ (~2.5 sccm) | 0.04 | 3 mins |

We note that under the growth conditions outlined above in the second row of Table 1, the resulting single crystalline BLG flakes do not favor any specific twist angle between layers. In contrast, for the growth parameters listed below in the second row of Table 2, most of the BLG flakes exhibit small twist-angle configurations. We repeat in the first row of Table 2 the same conditions shown in the second row of Table 1 for easy comparison of the growth conditions for BLG with random twist-angles (first row) and small twist-angle distributions

(second row).

| Table 2 | Plasma (W) | $P_{CH4}$ (torr) | $P_{H2}$ (torr) | $P_{CH4}/P_{H2}$ | Growth Time |
|---|---|---|---|---|---|
| BLG (random) | 40 | $7\times10^{-7}$ | $1.6\times10^{-5}$ (~ 2.5 sccm) | 0.04 | 3 mins |
| BLG (small-angle distributions) | 40 | $9\times10^{-7}$ | $1.75\times10^{-5}$ (~ 2.7 sccm) | 0.05 | 3 mins |

The interlayer twist angle has been determined through SEM imaging by measuring the neighboring edges of the first and second layers.

## 2) Sample preparation and growth parameters for large single crystalline BLG flakes

For the growth of large single crystalline BLG flakes, additional steps are needed in the preparation of the Cu substrates, as detailed below.

Cu-foils purchased from Nilacon with a thickness of 100 μm and a purity of 99.96% are prepared within following steps: *i*) sonication in ACE and IPA, *ii*) soaked in acetic acid ($CH_3COOH$) at 80 °C for 35 mins for the removal of $CuO_x$, and *iii*) annealed at 1050 °C in a mixture of Ar and $H_2$ gases for 30 mins to smooth the Cu surface. After the annealing process, the annealed Cu-foils are then inserted into PECVD system (Figure S1) for graphene growth. The growth parameters are as follows: plasma power = 40 W, $P_{CH4} = 7 \times 10^{-7}$ torr, $P_{H2} = 1.6 \times 10^{-5}$ torr (~ 2.5 sccm), and growth time = 20 mins. Under these growth parameters, the grain size of MLG and BLG can reach 15 μm and 5 μm, respectively. In this case, the interlayer twist angle in BLG does not favor any specific rotational configuration. Figure S2 shows the optical microscopic images of hexagonal single crystalline graphene grown on 100 μm thick Cu-foils and the corresponding Raman spectra.

## 3) Growth of large BLG films with small interlayer twist angles

Figure S3 (a)–(d) shows the SEM images of growth results with the evolution of $CH_4$-to-$H_2$ pressure ratio ($P_{CH4}/P_{H2}$) from 0.03 to 0.06 under a given $H_2$ flow rate at 3 sccm, and the partial pressures are determined by RGA and found to be ~ $2\times10^{-5}$ torr for $H_2$. The white regions indicate the optical contrast for bare 285 nm $SiO_2$/Si substrate. The light and dark gray areas correspond to MLG and BLG regions, respectively. The density of BLG increases with increasing ($P_{CH4}/P_{H2}$). According to the SEM images (Fig. S3(d)), it's clear to see that some multilayer graphene exhibited in the mm-size BLG. The growth parameter that we used in displaying Fig. S3(d) shows that most of BLG prefers to arrange in the small angle configuration in accordance with the analysis of Raman spectroscopy (Fig. S4).

**Supplementary Figure 2:**

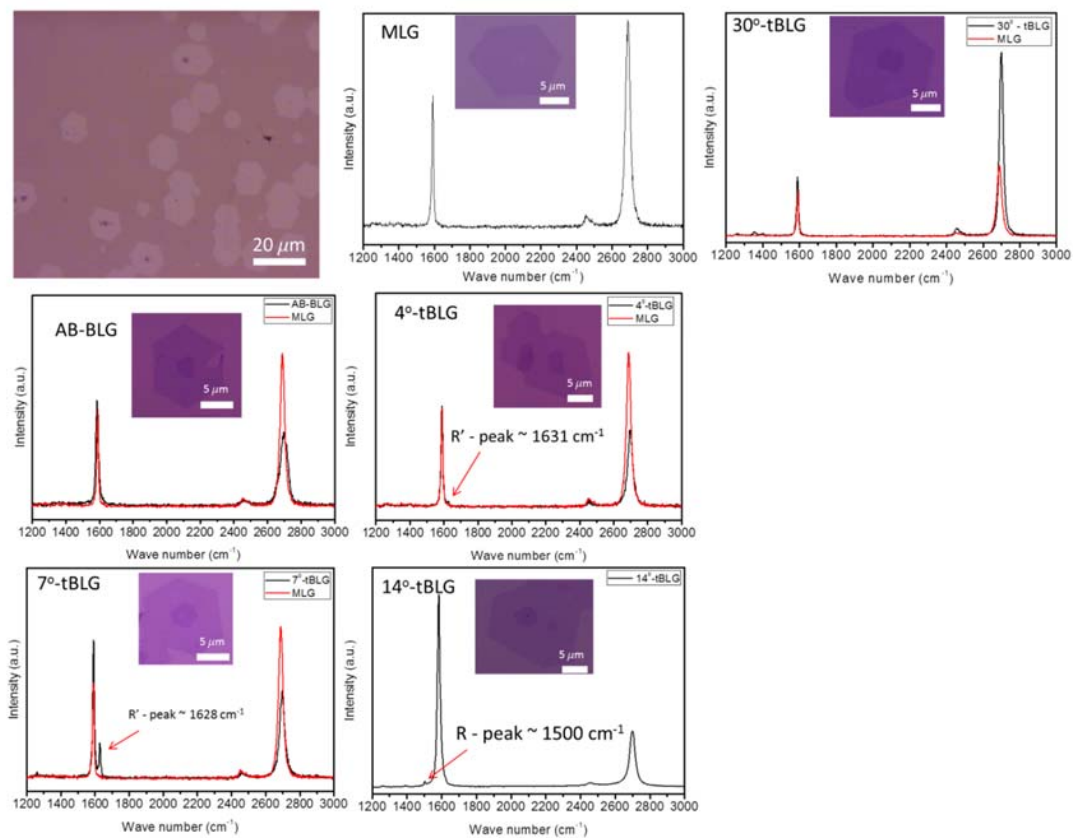

**Fig. S2** shows the single crystal MLG and BLG on Cu-foil, the second layer is hard to observe unless the Cu-foil is baked in order to reveal the color contrast. While, other images show the database of Raman spectrum in which the twisted angle was directly measured by the neighboring edges between layers. The related Raman mapping results are shown in Fig. 5 in the main text. Except the OM image for graphene grown on Cu-foil, the scale bars shown in other OM images indicate 5 μm.

**Supplementary Figure 3:**

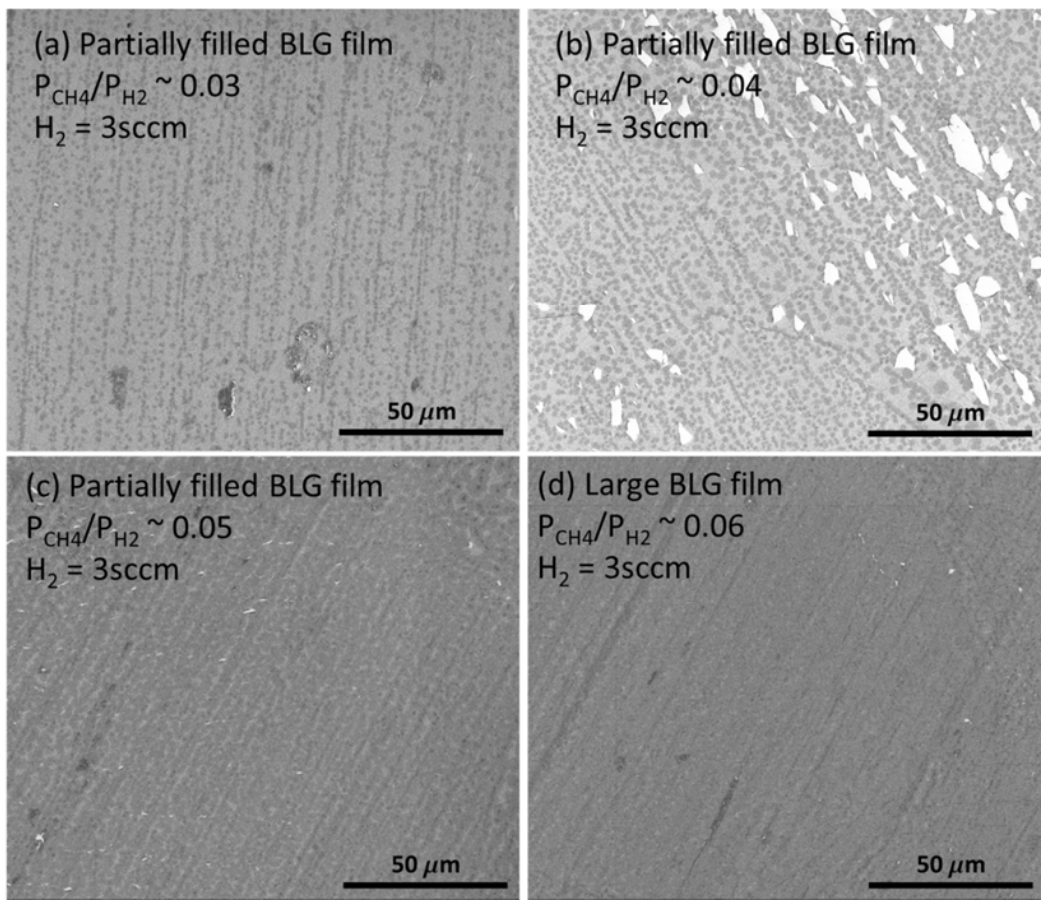

**Fig. S3.** SEM images showing the evolution of BLG growth with increasing $CH_4$-to-$H_2$ pressure ratio ($P_{CH4}/P_{H2}$): **(a)** ($P_{CH4}/P_{H2}$) = 0.03, **(b)** ($P_{CH4}/P_{H2}$) = 0.04, **(c)** ($P_{CH4}/P_{H2}$) = 0.05, **(d)** ($P_{CH4}/P_{H2}$) = 0.06.

**Supplementary Figure 4:**

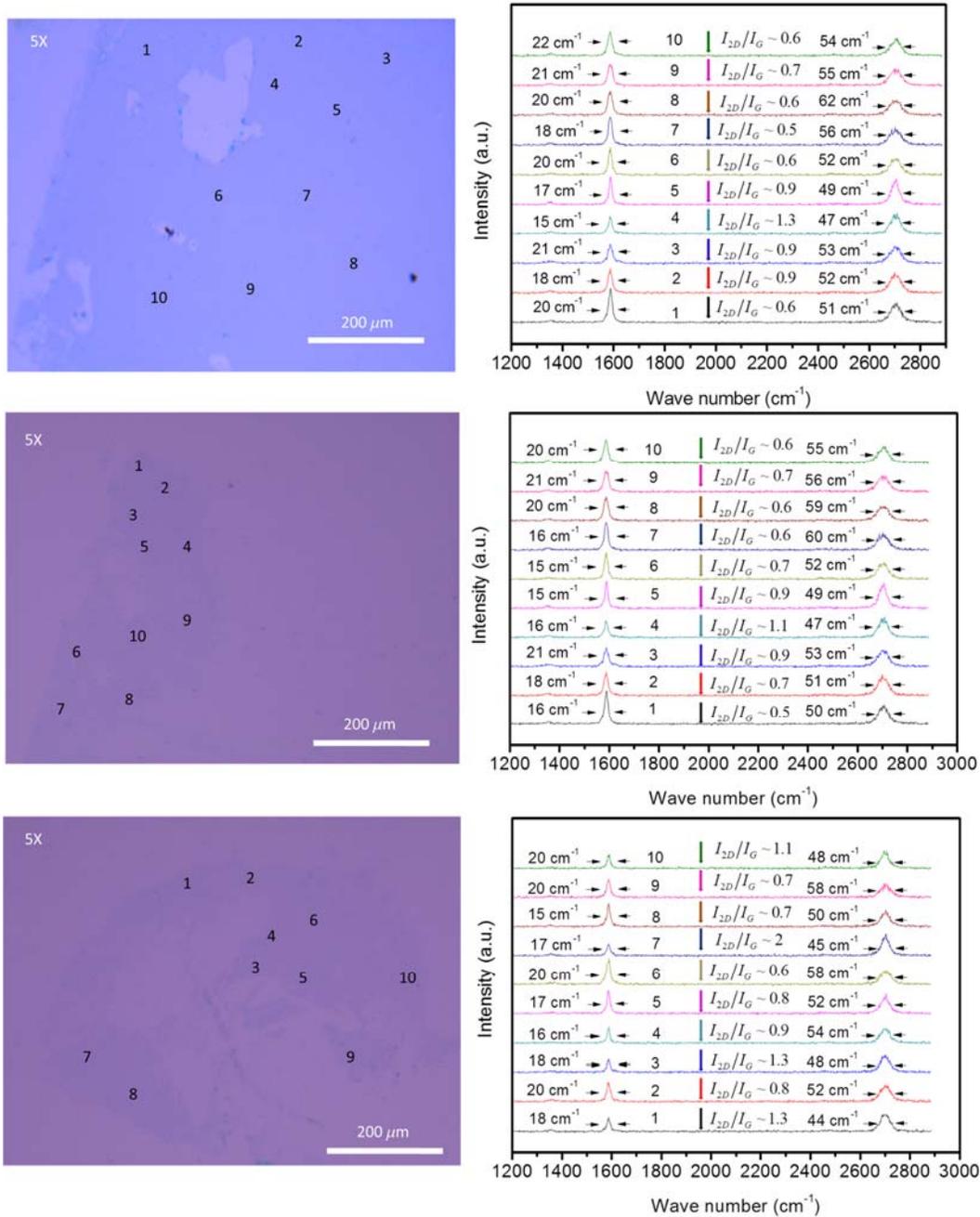

**Fig. S4.** SEM images and Raman point spectra taken on an mm-size uniform BLG film: We randomly select 20 different points over the BLG large film and check the corresponding Raman point spectra, with the first set of 10 points indicated on the SEM image (top left) and the corresponding Raman spectra shown in the top right panels, and the second set of 10 points indicated on the SEM image (bottom left) and corresponding Raman spectra shown in the bottom right panel. The position for each point Raman spectrum is first located by 5× optical magnification then the spectrum is taken under 100× magnification.

## 4) Correlation of the twist angle distributions with the partial pressures of CH$_4$ and H$_2$

In Figure S5 we summarize how the growth of tBLG evolve with the partial pressure ratio (P$_{CH4}$/P$_{H2}$), which plays a key role not only in the BLG growth but also in controlling the twist angle between layers of BLG. We note that the H$_2$ flow rate used for sizable BLG growth in our PECVD system ranged from 2.5 sccm (corresponding to P$_{H2}$ ~ 1.6×10$^{-5}$ torr) to 3 sccm (corresponding to P$_{H2}$ ~ 2×10$^{-5}$ torr). Within this limited range of H$_2$ flow rate, a larger H$_2$ flow rate would lead to a slightly larger partial pressure ratio (P$_{CH4}$/P$_{H2}$) necessary to control the twist angle but would not significantly alter the trend of the control of the twist angle, as detailed in Table S1 shown below.

| H$_2$ flow rate | (P$_{CH4}$/P$_{H2}$) for small twist angles | (P$_{CH4}$/P$_{H2}$) for arbitrary twist angles |
|---|---|---|
| 2.5 sccm | ~ 0.03 (P$_{CH4}$ ~ 5×10$^{-7}$ torr) | ~ 0.04 (P$_{CH4}$ ~ 7×10$^{-7}$ torr) |
| 2.7 sccm | ~ 0.05 (P$_{CH4}$ ~ 9×10$^{-7}$ torr) | ~ 0.07 (P$_{CH4}$ ~ 1.2×10$^{-6}$ torr) |
| 3 sccm | ~ 0.06 (P$_{CH4}$ ~ 1.2×10$^{-6}$ torr) | ~ 0.08 (P$_{CH4}$ ~ 1.6×10$^{-6}$ torr) |

**Table S1.** Summary of the correlation between the H$_2$ flow rate and the partial pressure ratio (P$_{CH4}$/P$_{H2}$) necessary for controlling the twist angle in tBLG.

To investigate the effect of (P$_{CH4}$/P$_{H2}$) on the twist angle between layers in tBLG, we chose a H$_2$ flow rate = 3 sccm, and found that when (P$_{CH4}$/P$_{H2}$) equalled either 0.06 or 0.08, the resulting large BLG films were uniform but revealed different twist-angle distribution for different (P$_{CH4}$/P$_{H2}$) values. As exemplified in Figs. S5 (a) and (b) for 10 representative Raman spectra taken on large BLG films synthesized with (P$_{CH4}$/P$_{H2}$) = 0.06 and 0.08, respectively, the smaller ratio (P$_{CH4}$/P$_{H2}$) = 0.06 yielded a preferably smaller angle distribution shown by the histogram in Fig. S5 (d), whereas the larger ratio (P$_{CH4}$/P$_{H2}$) = 0.08 resulted in arbitrary angle distribution shown in Fig. S5 (e). In contrast, for larger ratio (P$_{CH4}$/P$_{H2}$) = 0.12, the quality of tBLG film degraded significantly so that the uniformity of tBLG film became compromised, as manifested by the SEM images in Figs. S5 (g), (h) and (i) for (P$_{CH4}$/P$_{H2}$) = 0.06, 0.08 and 0.12, respectively, and also by the strong Raman D-peak revealed in Fig. S5 (c) for the BLG film grown with (P$_{CH4}$/P$_{H2}$) = 0.12. We further remark that the histograms shown in Figs. S5 (d), (e), and (f) for the angular distributions of large tBLG films with different (P$_{CH4}$/P$_{H2}$) values were obtained by analysing the twist-angle dependent Raman spectral characteristics on more than 50 points for each sample at random locations, and the spectral characteristics for identifying the twist angles included the appearance of R-, R′- and non-dispersive D-band, and the resonant enhancement of G-band.

**Supplementary Figure 5:**

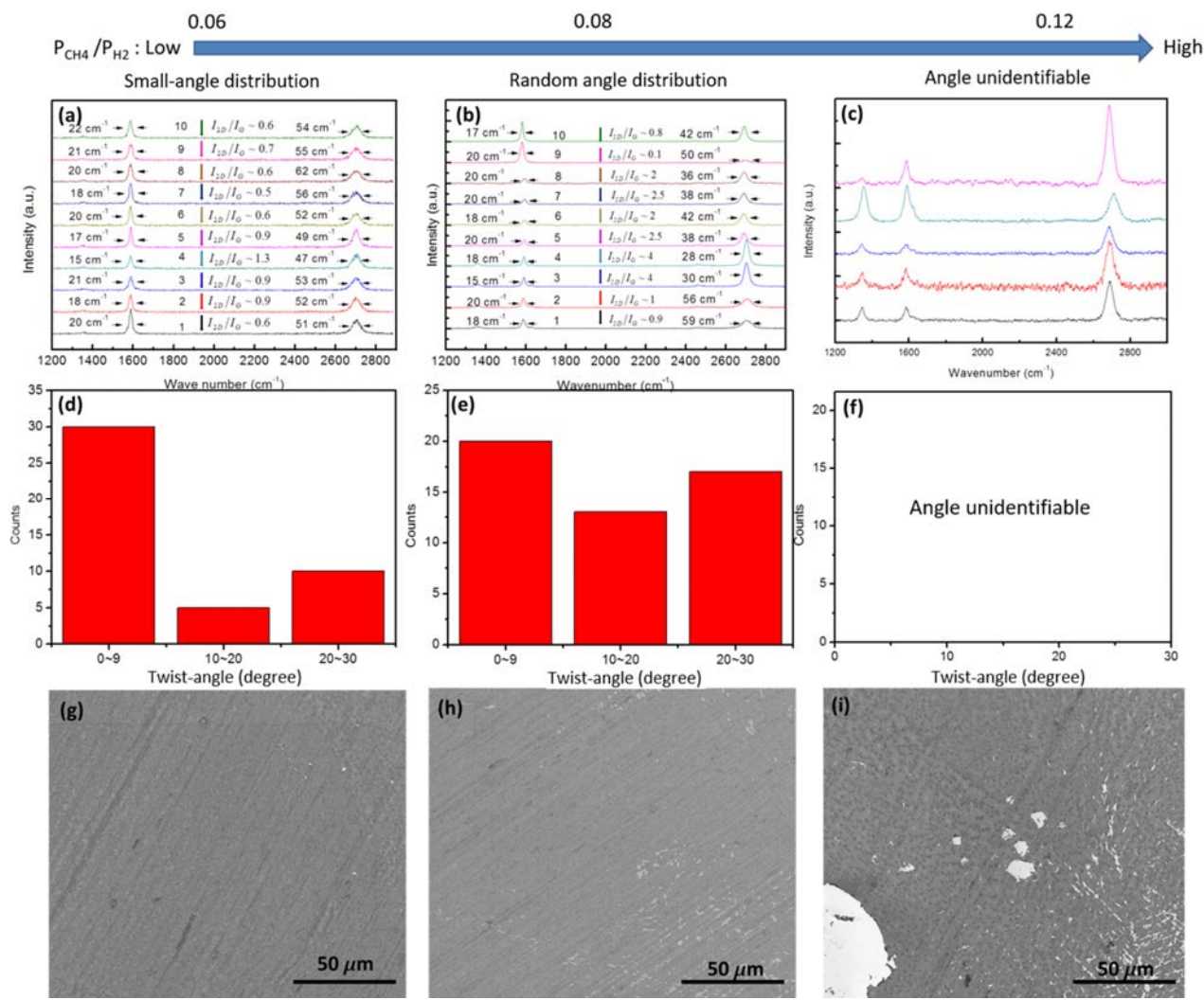

**Fig. S5.** Correlation of the partial pressure ratio ($P_{CH4}/P_{H2}$) with the twist angle distribution in PECVD-growth tBLG samples: **(a) – (c)** Point Raman spectra taken at 10 different locations on each mm-size tBLG sample for ($P_{CH4}/P_{H2}$) = 0.06, 0.08 and 0.12, respectively. **(d) – (f)** Histograms of the angular distribution determined from 50 random locations of each mm-size tBLG sample for ($P_{CH4}/P_{H2}$) = 0.06, 0.08 and 0.12, respectively. **(g) – (i)** Representative SEM images of the mm-size tBLG samples with ($P_{CH4}/P_{H2}$) ratio = 0.06, 0.08 and 0.12, respectively. The SEM image in (i) suggests serious nonuniformity of tBLG sample synthesized with the condition ($P_{CH4}/P_{H2}$) ≥ 0.12, and the corresponding Raman spectra in (c) do not exhibit recognizable signatures for identifying the twist angles in the BLG sample.

**Supplementary Figure 6:**

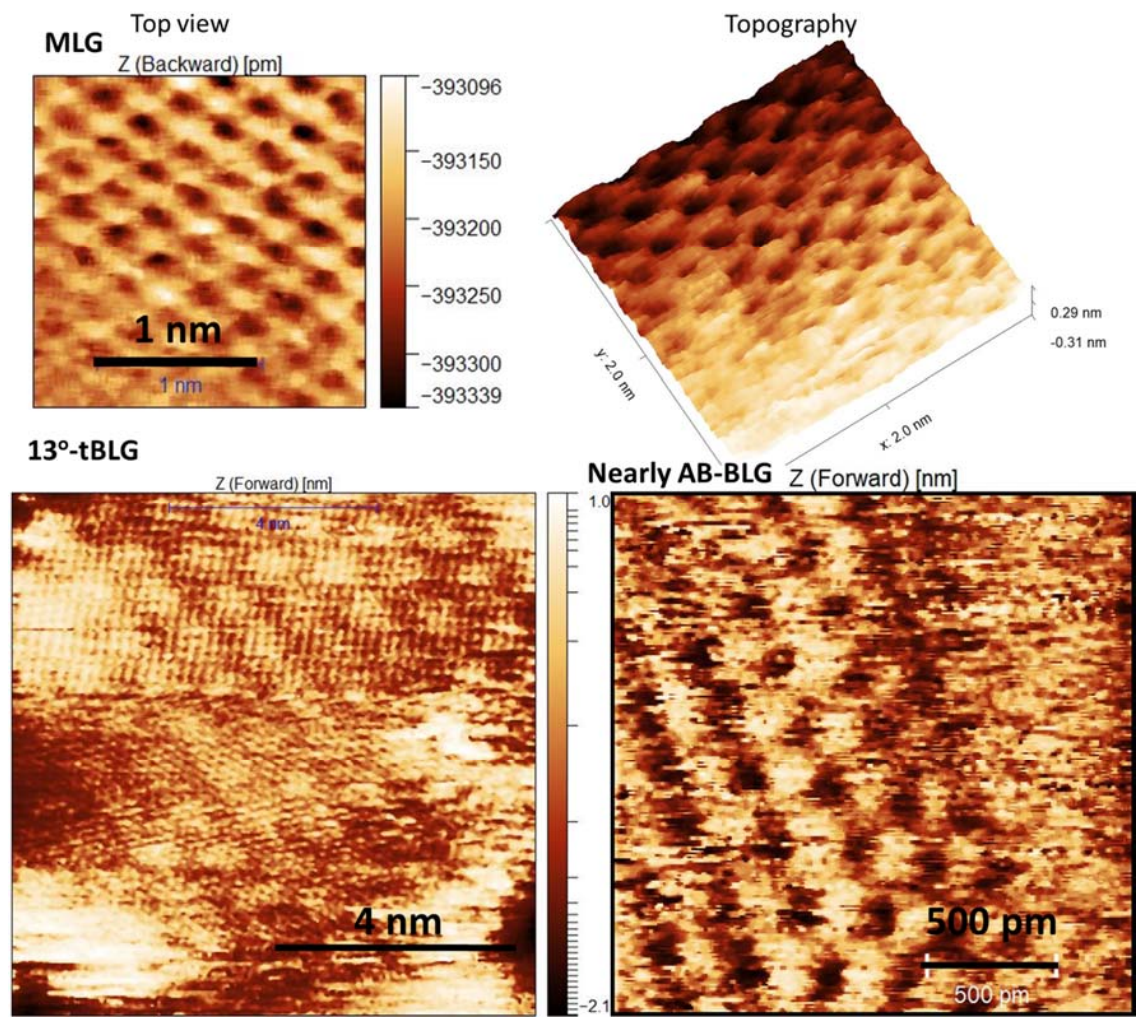

**Fig. S6.** Atomically resolved topographic images of MLG and tBLG taken by a scanning tunneling microscope: The top two images reveal the STM top view and topography for MLG, where the yellow scale bar indicates 1 nm. The bottom two images show the Moiré patterns of tBLG with the twist-angle $\theta = 13°$ (left panel) and nearly AB-BLG (right panel). The black scale bars in the left and right panels are 4 nm and 500 pm, respectively.